\documentclass[
reprint,
 amsmath,amssymb,
 aps,
pra,
]{revtex4-2}
\usepackage[utf8]{inputenc}
\usepackage[version=3]{mhchem}
\usepackage{amsmath}
\usepackage{gensymb}
\usepackage{color}
\usepackage{soul}
\usepackage{braket}
\usepackage{xr-hyper}
\usepackage{hyperref}
\usepackage{graphicx}
\usepackage{dcolumn}
\usepackage{bm}
\usepackage{tikz}
\usepackage{ragged2e}
\usepackage{kantlipsum}
 \usepackage{setspace}
 \usepackage{upgreek}
\usepackage{capt-of}
\usepackage{float}
\usepackage[normalem]{ulem}
\usetikzlibrary{matrix}
\usepackage{eqparbox}
\newbox\matrixcellbox
\tikzset{center align per column/.style={nodes={execute at begin
            node={\setbox\matrixcellbox=\hbox\bgroup},
            execute at end
            node={\egroup\eqmakebox[\tikzmatrixname\the\pgfmatrixcurrentcolumn][c]{\copy\matrixcellbox}}}},
}

\usepackage{xcolor}
\usepackage[draft,inline,nomargin,index]{fixme}

\fxsetup{theme=color,mode=multiuser}

\FXRegisterAuthor{lee}{alee}{\color{red} lee}
\FXRegisterAuthor{ofer}{aOfer}{\color{brown} Ofer} 
\FXRegisterAuthor{tomer}{atomer}{\color{magenta} Tomer} 
\FXRegisterAuthor{bankim}{abankim}{\color{orange} \bf Bankim}
\FXRegisterAuthor{sasha}{asasha}{\color{orange} \bf Sasha}
\FXRegisterAuthor{eilon}{aeilon}{\color{magenta} Eilon} 

\definecolor{cadmiumgreen}{rgb}{0.0, 0.68, 0.24}

\newcommand{\rb}{r_\mathrm{b}}
\newcommand{\OD}{\mathrm{OD}}
\newcommand{\ODb}{\mathrm{OD}_\mathrm{b}}
\newcommand {\rmi}{{i}}
\newcommand {\e}{{e}}

\newcommand {\MHz}{\mathrm{MHz}}
\newcommand{\Oc}{\Omega}
\newcommand{\Deltac}{\Delta_\mathrm{c}}

\begin{document}
\title{Quantum vortices of strongly interacting photons}
\author{Lee Drori}
\thanks{These two authors contributed equally}
\affiliation{Department of Physics of Complex Systems, Weizmann Institute of Science, Rehovot 7610001, Israel}
\author{Bankim Chandra Das}
\thanks{These two authors contributed equally}
\affiliation{Department of Physics of Complex Systems, Weizmann Institute of Science, Rehovot 7610001, Israel}
\author{Tomer Danino Zohar}
\affiliation{Department of Physics of Complex Systems, Weizmann Institute of Science, Rehovot 7610001, Israel}
\author{Gal Winer}
\affiliation{Department of Physics of Complex Systems, Weizmann Institute of Science, Rehovot 7610001, Israel}
\author{\linebreak Eilon Poem}
\affiliation{Department of Physics of Complex Systems, Weizmann Institute of Science, Rehovot 7610001, Israel}
\author{Alexander Poddubny}
\affiliation{Department of Physics of Complex Systems, Weizmann Institute of Science, Rehovot 7610001, Israel}
\author{Ofer Firstenberg}
\affiliation{Department of Physics of Complex Systems, Weizmann Institute of Science, Rehovot 7610001, Israel}
\begin{abstract}
Vortices are a hallmark of topologically nontrivial dynamics in nonlinear physics and arise in a huge variety of systems, from space~\cite{Pines1985} and atmosphere~\cite{Morton1966} to condensed matter~\cite{Abrikosov2004,Zeldov2022} and quantum gases~\cite{Ketterle2001,Lagoudakis2008,Roumpos2010,Klaus2022}. In optics, vortices manifest as phase twists of the electromagnetic field, commonly formed by the interaction of light and matter~\cite{Shen2019}. Formation of vortices by effective interaction of light with itself requires strong optical nonlinearity and has therefore been confined, until now, to the classical regime \cite{Tikhonenko1996,Mamaev1996}. Here we report on the realization of quantum vortices resulting from a strong  photon-photon interaction in a quantum nonlinear optical medium. The interaction causes faster phase accumulation for co-propagating photons. Similarly to a plate pushing water~\cite{Kiehn2001}, the local phase accumulation produces a quantum vortex-antivortex pair within the two-photon wavefunction.  
For three photons, the formation of vortex lines and a central vortex ring attests to a genuine three-photon interaction.
The wavefunction topology,  governed by two- and three-photon bound states, imposes a conditional phase shift of $\pi$-per-photon, a potential resource for deterministic quantum logic operations.

\end{abstract}
\maketitle

Photons essentially do not interact with one another in the optical regime. Effective interaction between photons can be mediated by matter, but, in conventional optical nonlinear media, this interaction is insignificant on the level of individual photons. It is only at the ultimate limit of quantum nonlinear optics, realized in specially-engineered systems, that a single photon can significantly alter the optical response of the system, rendering a meaningful photon-photon interaction~\cite{Imamoglu1997,Harris1999,Chang2014,Firstenberg2016}. 

The realization of strong photon-photon interactions in atomic ensembles has many consequences. It enables quantum-information processing for optical qubits, such as single-photon transistors and phase gates \cite{Kimble1995,Gorniaczyk2014,DurrPRX2022}. It can be used to generate non-classical states of light, such as squeezed, cluster, and repeater states, as a resource for quantum communication, computation, and sensing \cite{RauschenbeutelNatPhys2021,PanNatPhot2022}. And, most fascinatingly, it manifests interacting quantum gases and fluids of photons, leading the way to exotic many-body physics for light \cite{Carusotto2013,OtterbachPRL2013,GorshkovPRA2015,Gorshkov2017}.

\begin{figure*}
\centering\includegraphics[width=0.84\textwidth,trim={0cm 3cm 0 3cm},clip]{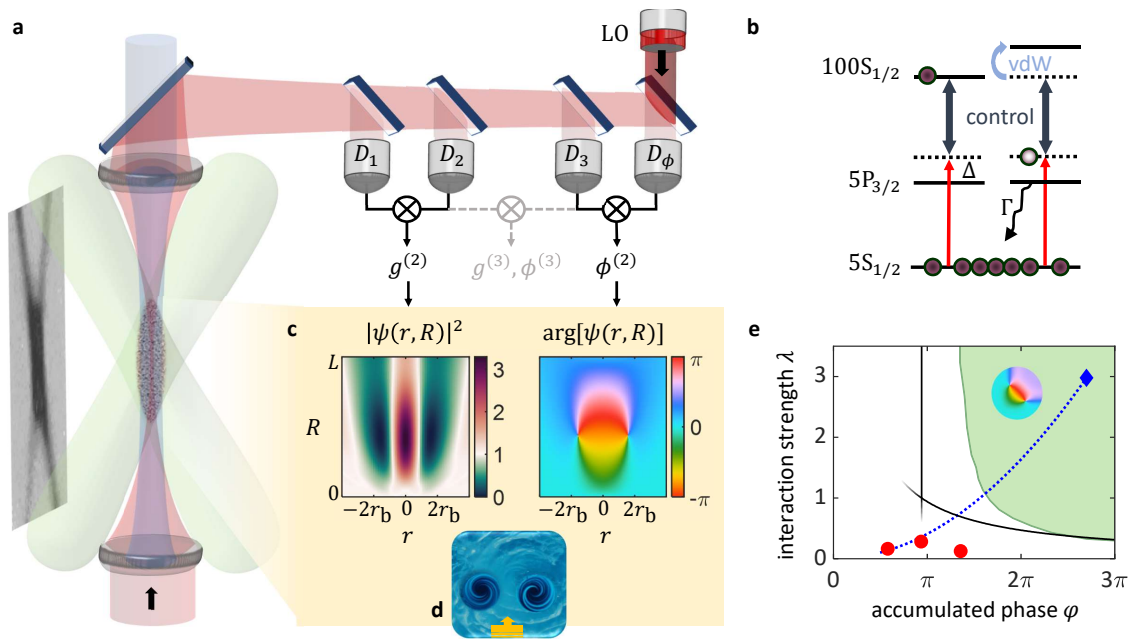}
\caption{{\bf Setup and conditions for generating photon vortices in Rydberg-mediated quantum nonlinear optics.} 
\textbf{a,} Counter-propagating probe (red) and control (blue) fields are focused onto an elongated ultracold atomic cloud (left, absorption image), held in a crossed optical trap (green). 
After traversing the medium, the probe light is split and measured by four single-photon detectors $D_{1-3,\phi}$. Detectors $D_{1,2,3}$ provide the two-photon and three-photon correlation functions $g^{(2)}$ and $g^{(3)}$. Detector $D_\phi$, measuring the interference beat between the probe and  reference light (LO: local oscillator) is correlated with detectors $D_{1,2,3}$ to extract the conditional phases $\phi^{(2)}$ and $\phi^{(3)}$ (see also Extended Data Fig.~\ref{fig:ED1}).
\textbf{b,} Atomic level-structure for generating polaritons comprising the $100S$ Rydberg orbital. The van-der-Waals interaction (vdW) between two Rydberg atoms disturbs the propagation of two close polaritons, leading to local accumulation of excess propagation phase.
\textbf{c,} Amplitude and phase of the two-photon wavefunction, calculated numerically in the Schr\"odinger approximation [Eq.~(\ref{eq:Schroedinger}), with $U=\sqrt{12}/\rb$], showing the formation of a vortex-antivortex pair of two interacting photons ($\rb$ is the blockade radius, $L$ is the length of the medium). \textbf{d,} The process resembles that of a vortex-antivortex pair forming in a water pool when a plate is pushed through the surface.
\textbf{e,} 
Conditions for vortex generation (shaded area, calculated numerically for a realistic Gaussian cloud) in terms of the photon-photon interaction strength $\lambda$ and the accumulated interaction phase in the finite medium $\varphi$. 
The solid curves are minimal conditions calculated analytically for a uniform cloud under the approximation of Schr\"odinger evolution. The dotted blue line indicates the available $\varphi$ and $\lambda$ in our experimental setup, peaking at the blue diamond. Red circles indicate the conditions of Refs.~\cite{firstenberg2013attractive,Porto2021,Liang2018}.
}\label{fig:1}
\end{figure*}

In this work, we realize an extreme regime of quantum nonlinear optics and observe, for the first time, optical quantum vortices generated by the interaction between only two or three photons. Quantum vortices -- phase singularities of the wavefunction -- are typically expected in strongly-interacting systems of many particles, primarily superfluids~\cite{Gauthier2019,Johnstone2019,Klaus2022}. Here we induce a strong, effective interaction at the few-photon level by coupling photons to Rydberg atoms in an ultracold rubidium gas, as sketched in Fig.~\ref{fig:1}a,b. 

The photons propagate through the medium as light-matter polaritons, at $10^{-6}$ of the speed of light. Due to the strong van-der-Waals coupling between the Rydberg atoms, photons that propagate close to each other -- at a distance smaller than the so-called blockade radius $\rb$ -- experience a local change in the refractive index. For two photons, the optical field can be described by a wavefunction $\psi(x_1,x_2)$, which is the probability amplitude of having photons at coordinates $x_1$ and $x_2$ along the medium \cite{firstenberg2013attractive}. The change in refractive index at $|x_1-x_2|\le \rb$ causes accumulation of excess local phase in $\psi(x_1,x_2)$. Remarkably, given sufficient conditions, this phase accumulation produces vortices in $\psi(x_1,x_2)$.

To understand the vortex formation, we consider a simplified analytical model describing the evolution of $\psi(x_1,x_2)$ in a medium of uniform density and length $L$ using the Schr\"odinger equation \cite{firstenberg2013attractive,FleischhauerPRA2017}
\begin{equation}\label{eq:Schroedinger}
i \partial_{R} \psi=-\frac{1}{2m } \partial_{r}^2 \psi +V^{(2)}(r)\psi. 
\end{equation}
Here $r=x_1-x_2$ is the relative coordinate of the two photons, and the center-of-mass coordinate $R=(x_1+x_2)/2=0\ldots L$ along the medium plays the role of time. The incoming, coherent probe field dictates homogeneous initial conditions $\psi(r,R=0)=1$. The photon-photon interaction is described by a nearly-square potential $V^{(2)}(r)=U/(1+r^6/\rb^6)$, where $U$ constitutes the change in the refraction index due to the Rydberg blockade, and the (negative) effective mass $m=-U/8$ arises from single-photon dispersion. Since $mU<0$, Eq.~(\ref{eq:Schroedinger}) describes attraction between photons \cite{firstenberg2013attractive}.

The striking and nontrivial prediction of Eq.~\eqref{eq:Schroedinger} is the formation of vortex-antivortex pairs in the medium, symmetrically around the potential well. This is demonstrated in Fig.~\ref{fig:1}c by a rigorous solution of Eq.~\eqref{eq:Schroedinger}. Equation \eqref{eq:Schroedinger} has localized and extended eigenstates, denoted as bound and scattering two-photon states \cite{firstenberg2013attractive}. Considering a single bound-state $\psi_{\rm bound}(r)$ with energy $E_2$, which is the case in our experiment, the initial scattering component is $\psi_\mathrm{scat}=1-\psi_{\rm bound}(r)$.
In the crudest approximation, the phase accumulation of the scattering component can be neglected (see SI), and the solution to Eq.~\eqref{eq:Schroedinger} reduces to
\begin{equation}\label{eq:simple2}
\psi(r,R)=e^{-\rmi E_2R}\psi_{\rm bound}(r)+\psi_\mathrm{scat}.
\end{equation}
The phase vortices are formed around $\psi=0$, when the two terms -- the bound and scattering components -- cancel each other. This result is generic and occurs for a sufficiently long medium and for any number of bound states. 

Such interference mechanism for vortex-antivortex pair formation is universal~\cite{Berry1974} and known from acoustics~\cite{Berry1974} and atomic collisions~\cite{Hirschfelder1976} to linear \cite{Bliokh2012,Bliokh2021} and nonlinear \cite{Jhaij2016} optics.
It can even be reproduced at home by pushing a plate through water~\cite{Kiehn2001,PhysicsGirl2014}, as illustrated in Fig.~\ref{fig:1}d. However, it has not been observed before in quantum optics, because the required strong and prolonged photon-photon interaction was, thus far, unattainable. 

We quantify the duration and strength of the interaction by the dimensionless parameters $\varphi=UL/2$ and $\lambda=2 |U m| \rb^2$ \cite{Bienias2016}, respectively.  
For vortices to develop, $\varphi\lambda$ should be large. When the interaction is weak, \textit{i.e.} for $\lambda \ll 1$, we find the condition $\varphi\lambda>\varphi_0$, with the threshold phase $\varphi_0\approx 0.94\pi$. For moderate interactions $\lambda \gtrsim 1$, the condition is $\varphi  > \varphi_0$ (see SI). 
We illustrate these conditions by solid lines in the `phase diagram' in Fig.~\ref{fig:1}e, and they compare reasonably well with the domain of vortex formation calculated numerically for our experimental, nonuniform medium (green shading). 

In Rydberg polariton systems, the essential experimental parameters are the total optical depth $\OD$ of the medium and the optical depth of the blockade range $\ODb=\OD\cdot\rb/L$. These parameters govern $\varphi\propto\OD$ and $\lambda\propto\ODb^2$. In our experiment, we reach $\OD=110$ and $\ODb=22$, setting $\varphi\approx 2.7\pi$ and $\lambda\approx 3$ (blue diamond in Fig.~\ref{fig:1}e). The moderate interaction obtained in previous experiments ($\varphi\approx 1.5\pi$ and $\lambda\ll 1$), albeit being enough to observe the signatures of photonic bound states \cite{firstenberg2013attractive,Liang2018,Porto2021}, were well below the threshold for vortex formation.

\section*{Observation of two-photon vortices}
Our experiment starts by compressing and trapping an ultracold rubidium cloud with a Gaussian density profile $\rho_0 e^{-x^2/2\sigma^2}$ and effective length $L=\sqrt{2\pi}\sigma=75~\upmu$m along the optical axis $x$. Our maximal peak density $\rho_0=5\cdot10^{12}$~atoms/cm$^3$ is 3$-$6 times higher than in previous experiments and is the primary source of our large $\lambda\propto\rho_0^2$. We form Rydberg polaritons using counter-propagating probe and control fields, which together resonantly couple the atomic $5S$ ground level to the $100S$ Rydberg level via the $5P$ level, see Fig.~\ref{fig:1}b. We send on average $f=0.25$~probe photons per $\upmu$s.

The probe photons experience the three-level optical susceptibility, except within the blockade range ${\rb=15.3~\upmu}$m, where they experience the two-level ($5S$-$5P$) susceptibility. Ideally, we desire the difference $U$ between these susceptibilities to be purely real, rendering a conservative photon-photon interaction. To this end, we first detune the probe from the atomic transition by 
$\Delta/(2\Gamma)=4.7$ linewidths and then adjust the control frequency such that the transmission remains the same outside and inside of the blockade range \cite{firstenberg2013attractive}. This adjustment eliminates the residual non-conserving (dissipative) part of the interaction and amplifies the conserving part $U=(\OD/L)(q\Gamma/\Delta)$ by the factor $q\approx1.4$ (see Methods). 
The atomic density declines during the experimental cycle, allowing us to study the full range $110 \ge\OD \ge 20$ ($2.7\pi \ge \varphi \ge 0.5\pi$) and correspondingly $22 \ge\ODb \ge 4.1$ ($3 \ge \lambda \ge 0.1$) in each experiment (dotted line in Fig.~\ref{fig:1}e).

The vortex-antivortex pair forms in the two-photon wavefunction $\psi(r,R)$ of the probe when it traverses the medium. While we cannot access these vortices directly inside the medium, we can observe them at the medium's boundary by measuring the dependence of $\psi(r,R=L)$ on the optical depth $\OD$. The larger the $\OD$, the longer the effective interaction between the two photons, so the dependence on $\OD$ at the boundary becomes a proxy for the evolution in the bulk. 
While $\OD$ corresponds to $R$, the temporal separation $\tau$ between the outgoing photons coarsely corresponds to their spatial separation $r$ in the medium \cite{Peyronel2012}. Experimentally, we measure the two-photon correlation function $g^{(2)}(\tau)$ and conditional phase $\phi^{(2)}(\tau)$ of the outgoing probe field for varying $\OD$ and associate them, respectively, with the squared amplitude and phase of $\psi(r,R=L)$. 

\begin{figure}[t!]
\centering
\includegraphics[width=0.93\columnwidth,trim={0cm 0 0.5cm 0cm},clip]{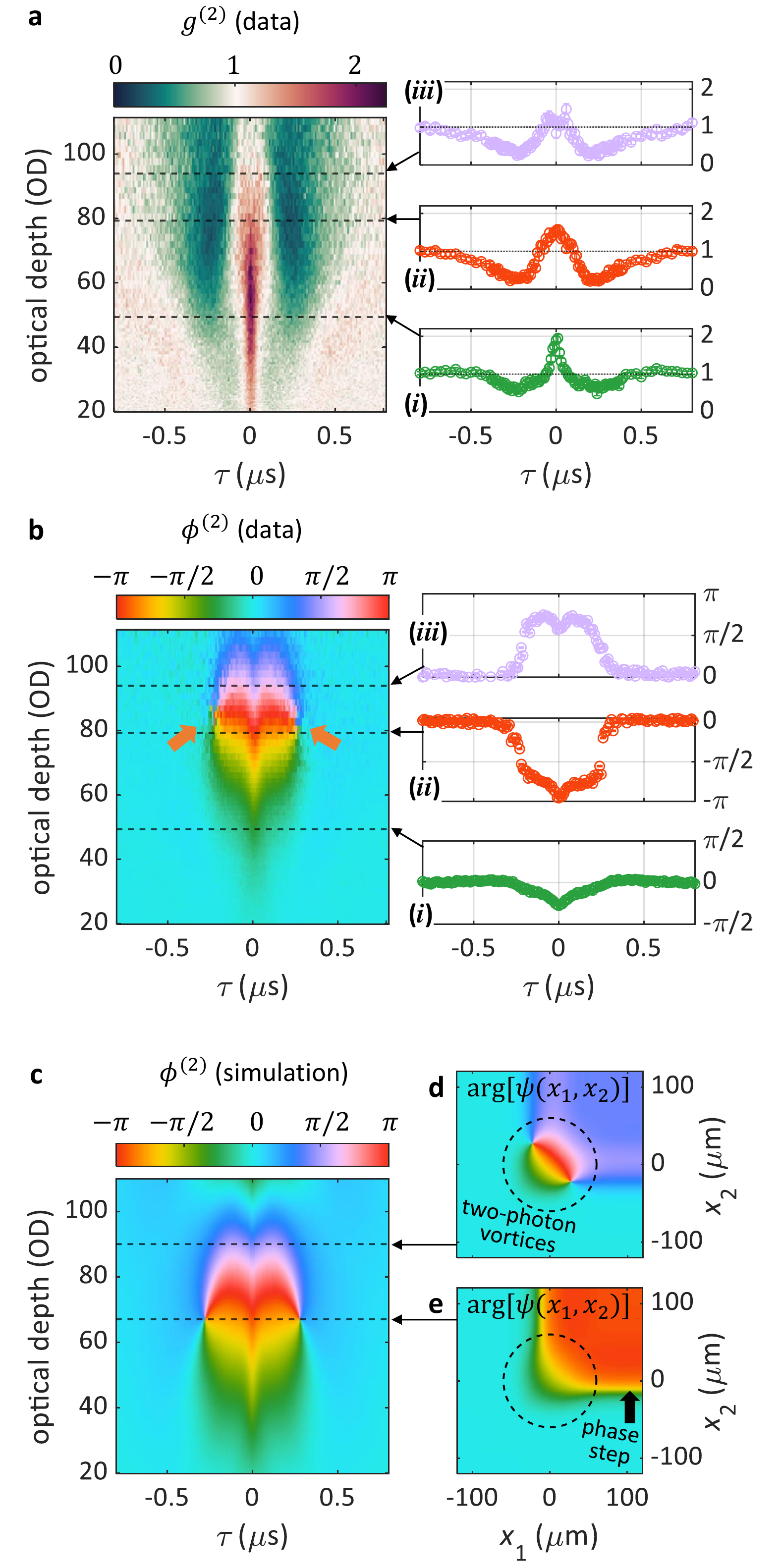}
\caption{ {\bf Two-photon vortices.} \textbf{a,b, } Color maps show the experimental (\textbf{a}) two-photon correlation $g^{(2)}$ and (\textbf{b}) two-photon phase $\phi^{(2)}$ as a function of the time between photons $\tau$ and the optical depth of the atomic medium OD. Insets (\textit{i})-(\textit{iii}) explicitly show the measured curves $g^{(2)}(\tau)$ and $\phi^{(2)}(\tau)$ for $\OD=49, 78, 95$. A vortex-antivortex pair (orange arrows) is formed around $\OD=80$ on the edge of the medium and is thus captured by the detectors as a phase step around $\tau=0.25~\upmu\text{s}$ and a phase wrap of $\phi^{(2)}(\tau=0)$ from $-\pi$ to $+\pi$. At the vortices' core, $g^{(2)}$ approaches zero. 
\textbf{c}, Numerically calculated $\phi^{(2)}$ for the experimental conditions.
\textbf{d,e}, Numerically calculated phase of the stationary, space-dependent two-photon wavefunction, showing the two-photon vortices forming \textbf{(d)} deep inside the medium and \textbf{(e)} on the edge. Dashed circles represent the edge ($2\sigma$) of the Gaussian cloud.
}
\label{fig:vortex}
\end{figure}

The experimental results are shown in Fig.~\ref{fig:vortex}a,b.
We first observe the gradual bunching of photons, $g^{(2)}(0)>1$, as the OD increases, accompanied by depletion of $g^{(2)}(\tau)$ at $\tau\approx\pm 0.25~\upmu\text{s}$ [see cross-section (a.\textit{i})]. The bunching and depletion are due to the effective attraction between the photons, governed by a two-photon bound state and accompanied by accumulation of conditional phase $\phi^{(2)}(0)<0$ [see (b.\textit{i})] \cite{firstenberg2013attractive}. At $\OD=78$, $\phi^{(2)}(0)$ reaches $-\pi$. A $\pi$ conditional phase allows for deterministic, maximally-entangling operation on photonic qubits \cite{DurrPRX2022}, and here it is obtained for co-propagating photons for the first time.

\begin{figure*}[t!]
\centering\includegraphics[width=1.0\textwidth, trim={3cm 7.5cm 7cm 0},clip]{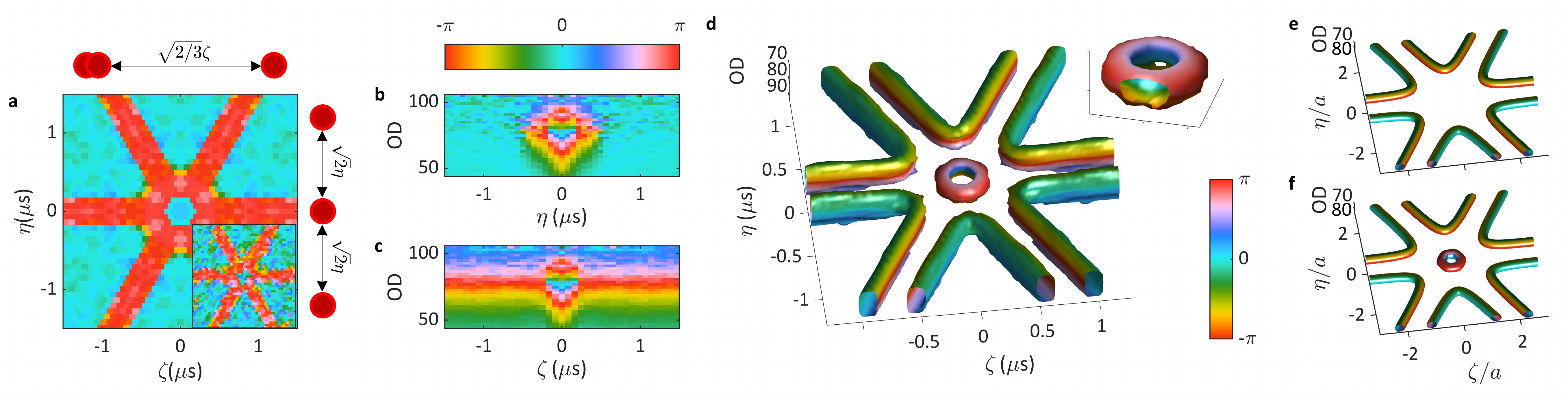}
\caption{\textbf{Three-photon vortex lines and vortex ring.} \textbf{(a-d)} Three-photon conditional phase $\phi^{(3)}(\eta,\zeta)$ measured at different ODs, where $\eta$ and $\zeta$ are time separations between the photons in Jacobi coordinates. For each OD, the data are averaged by assuming a six-fold symmetry; an example of data before averaging is shown in \textbf{a} (inset).
We present three cross-sections: \textbf{(a)} $\OD=79$, \textbf{(b)} $\zeta=0$, \textbf{(c)} $\eta=0$. In all these, the central feature originates from three-photon interactions. \textbf{(d)} To reveal the vortex structure, we plot the measured $\phi^{(3)}$ along isosurfaces of $|\nabla\phi^{(3)}|$, which are themselves derived from the data. We identify six vortex lines (tubes) and a central vortex ring (torus); note the $2\pi$ phase-twist around the cross-section of the tubes and (inset) torus. \textbf{(e,f)} Analytic calculation of the three-photon phase $\mathrm{arg}[\psi(\upeta,\upzeta,R)]$ using the approximate ansatz in Eq.~(\ref{eq:simple4}) without (e) and with (f) the three-photon bound-state term, for $E_3=3E_2$ and $\psi^{(3)}_{\rm bound}(0,0)=-3\psi^{(2\rightarrow3)}_{\rm bound}(0,0)$. The outer vortex lines are due to the generalized two-photon bound states, whereas the vortex ring is due to the three-photon bound state.
}

\label{fig:3}
\end{figure*}

The vortex-antivortex pair is formed around $\OD=80$, as manifested by the clockwise and anti-clockwise phase twists in Fig.~\ref{fig:vortex}b (orange arrows). These phase twists correspond to steep steps of $\phi^{(2)}(\tau)$ at $\tau=\pm 0.25~\upmu\text{s}$, whose direction flips when the OD increases [compare (b.\textit{ii}) to (b.\textit{iii})]. Furthermore, as expected, the photons are depleted from the vortex core, where $g^{(2)}(\tau)=0.24$ (global minima in Fig.~\ref{fig:vortex}a).
Following the vortex-antivortex formation, $\phi^{(2)}(0)$ decreases gradually from $\pi$ to 0, and the bunching and depletion in $g^{(2)}$ get smaller. The vortices effectively `unwind the tension' between the regions of fast and slow accumulation of phase.

Our experimental results are also corroborated by rigorous numerical simulations based on the model from Refs.~\cite{Peyronel2012,firstenberg2013attractive}. The simulations use the actual experimental parameters, including the Gaussian density profile of the medium and the propagation of the second photon after the detection of the first photon. They are presented in Fig.~\ref{fig:vortex}c and are in excellent agreement with the measured data, reproducing the vortex-antivortex pair and the surrounding phase structure in full detail. The 10\% discrepancy in the OD at which the vortices are observed is due to a slight saturation of our phase detection scheme and is completely resolved for a weaker incoming probe (see Fig.~SI.3).
We use these simulations to calculate the conditions for vortex formation in Fig.~\ref{fig:1}e (shaded area). 

Moreover, the simulations are able not only to reproduce the experimentally accessible data but also to establish the dynamics of the two-photon wavefunction $\psi(x_1,x_2)$ inside the medium. Figures \ref{fig:vortex}e,d present the phase of $\psi(x_1,x_2)$ for two values of $\OD$. These simulations exhibit the entrance of phase singularities and eventually the presence of a stationary pair of quantum vortices in the medium.

\section*{Three-photon vortices}

The strong interaction, responsible for forming two-photon vortex-antivortex pairs, leads to an even richer topological structure of the three-photon wavefunction $\psi(x_1,x_2,x_3)$. We will first illustrate this by considering only the pairwise photon interactions. If only two of the three photons are close together, they attract each other and form a quasibound two-photon state. As the third photon remains unbound, the three pairs of vortices resulting from each of the three possible quasibound states would manifest as six vortex lines in the $(x_1,x_2,x_3)$ space.

To verify this prediction, we have measured the three-photon phase $\phi^{(3)}(t_1,t_2,t_3)$ and the corresponding correlation function $g^{(3)}(t_1,t_2,t_3)$ over a range of OD. While the stationary two-photon correlation function depends on a single time separation $t_1-t_2$, characterization of the three-photon wave function requires two time separations. These are conveniently parametrized by the Jacobi coordinates $\eta=t_{21}/\sqrt{2}$ and $\zeta=(t_{13}+t_{23})/\sqrt{6}$, where $t_{ij}=t_i-t_j$ \cite{Jachymski2016}.
The three-photon measurements are then given by three-dimensional datasets for  $\phi^{(3)}$ and $g^{(3)}$ that depend on $\eta$, $\zeta$, and OD. 

Figure \ref{fig:3}a shows $\phi^{(3)}(\eta,\zeta)$ at the critical OD where vortices appear. We identify cores of vortex lines with $\pi$-phase dislocations: along the edges of the six-bar star ($|\phi^{(3)}|=\pi$ region) and around the center ($\phi^{(3)}=0$ region). Figures \ref{fig:3}b,c show the development of the phase along the lines $\zeta=0$ (uniformly-separated photons) and $\eta=0$ (a photon at some distance from a pair). 
Notably, $\phi^{(3)}(0,0)$ varies monotonically with OD much faster than $\phi^{(2)}(0)$ does, as seen by comparing Fig.~\ref{fig:3}b to Fig.~\ref{fig:vortex}b. 

Eventually, the vortices structure is best understood from Fig.~\ref{fig:3}d, showing the reconstructed three-dimensional isosurfaces $|\nabla\phi^{(3)}|=0.7~\mathrm{rad/\upmu s}$, with the phase $\phi^{(3)}$ indicated by the surface color. 
The star-like structure of tubes in Fig.~\ref{fig:3}d corresponds to six vortex lines, in agreement with our prediction. We can additionally verify that this structure is due only to the two-photon attraction by constructing it from the measured two-photon data. This construction, shown in Extended Data Fig.~\ref{fig:ED2}, exhibits a star-like vortex structure nearly identical to that found in the three-photon data. 

The vortex star structure can be explained analytically by generalizing Eq.~\eqref{eq:simple2} to three photons. To this end, we replace $\psi_{\rm bound}(r)$ by a sum 
$\psi^{(2\rightarrow3)}_{\rm bound}=\psi_{\rm bound}(r_{21})+\psi_{\rm bound}(r_{13})+\psi_{\rm bound}(r_{32})$ of three quasibound states ($r_{ij}=x_i-x_j$), respecting the bosonic (permutation) symmetry of the three-photon wave function.
For simplicity, we assume a weakly-bound state $\psi_{\rm bound}(r)= 2 e^{-2|r|/a}$ in a delta potential, where $a\approx \rb/\lambda$ is the scattering length \cite{firstenberg2013attractive, BieniasPRA2014}. As shown in Fig.~\ref{fig:3}e using the spatial Jacobi coordinates $\upeta=r_{21}/\sqrt{2}$ and $\upzeta=(r_{13}+r_{23})/\sqrt{6}$, the generalized Eq.~(\ref{eq:simple2}) gives the six vortex lines. 
These results support our interpretation that the vortex lines in the three-photon wave function are a direct generalization of the two-photon vortices.

The full three-photon data, however, show richer behavior, which cannot be reduced to just independent pairwise photon attraction. The effect of a third photon approaching a bound pair is reflected in the central feature in Fig.~\ref{fig:3}c and, more clearly, as a torus around $\eta=\zeta=0$ in Fig.~\ref{fig:3}d, manifesting a vortex ring. This inner ring does not originate from two-photon quasibound states but rather from a genuine three-photon bound state $\psi_{\rm bound}^{(3)}$ \cite{Jachymski2016,Liang2018} that interferes with the scattering states. 

\begin{figure}
\centering\includegraphics[width=\columnwidth, trim={1.5cm 1cm 1cm 1.5cm},clip]{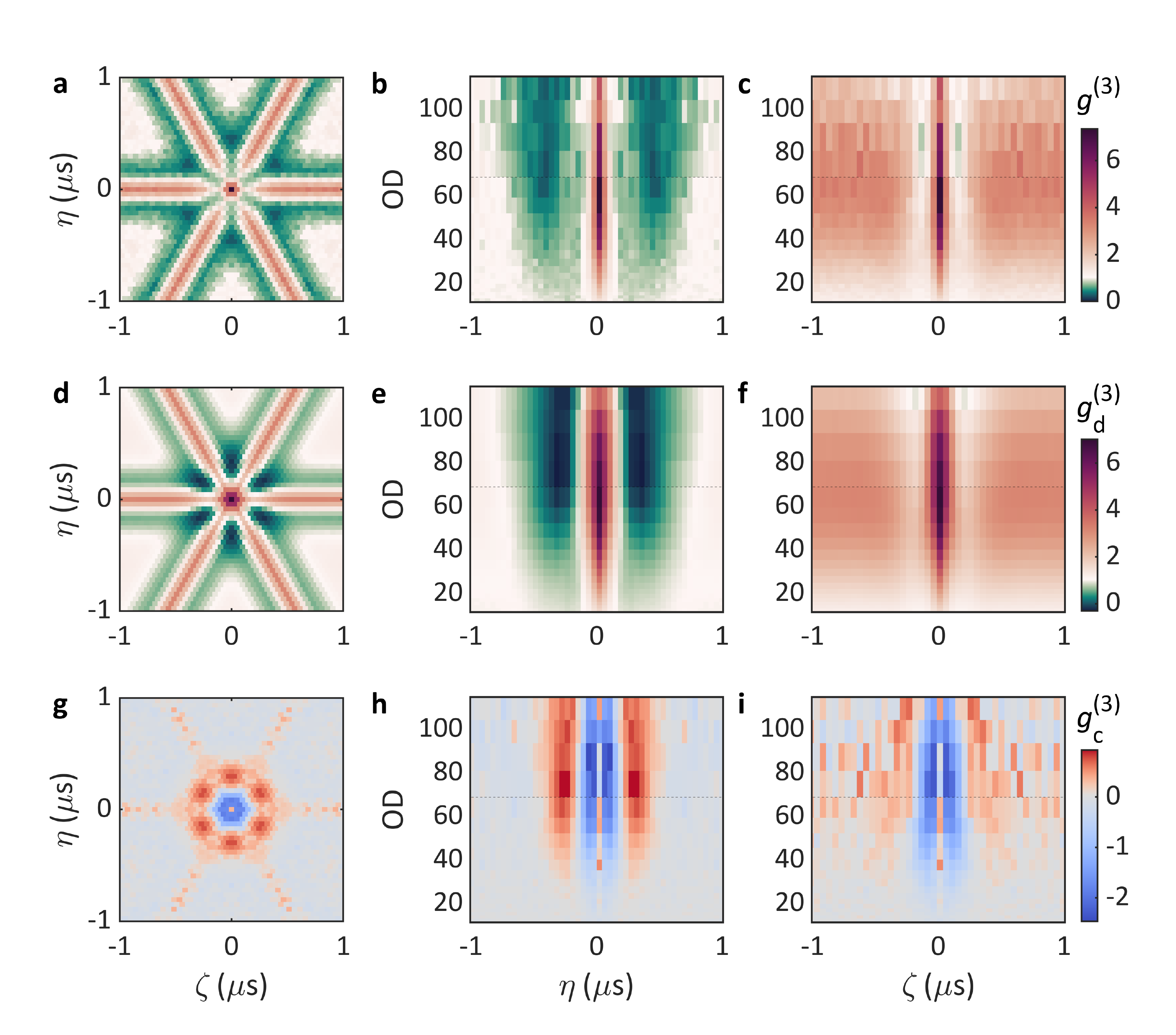}
\caption{{\bf Three-photon bound-state and blockade saturation.} \textbf{(a-c)} Three-photon correlations $g^{(3)}(\eta,\zeta,\OD)$ (after six-fold averaging), sectioned at \textbf{(a)} $\OD=69$, \textbf{(b)} $\zeta=0$, and \textbf{(c)} $\eta=0$. \textbf{(d-f)} Disconnected part of the correlations $g^{(3)}_\mathrm{d}(\eta,\zeta,\OD)$, calculated from only the $g^{(2)}(\tau,\OD)$ data and representing the two-photon contribution to the three-photon correlations. The strong confinement of the three-photon bound state leads to a tighter bunching feature. \textbf{(g-i)} Connected part of the correlations $g^{(3)}_\mathrm{c}=g^{(3)}-g^{(3)}_\mathrm{d}$, reflecting the dynamics beyond pairwise interactions. These dynamics are regulated by the saturation of the blockade interaction, expressed as an effective, short-range repulsive force. The negative central feature (blue), the positive ring (red), and the modulation along this ring are all governed by this force.
}
\label{fig:4}
\end{figure}

To describe the formation of the ring, we employ the following ansatz, adding $\psi_{\rm bound}^{(3)}$ and its energy $E_3$ to the generalized Eq.~\eqref{eq:simple2},
\begin{equation}\label{eq:simple4}
\psi(\upeta,\upzeta,R)=e^{-\rmi E_3R}\psi^{(3)}_{\rm bound}(\upeta,\upzeta)+e^{-\rmi E_2R}\psi^{(2\rightarrow3)}_{\rm bound}+\psi_\mathrm{scat},
\end{equation}
where $\psi_\mathrm{scat}=1-\psi_{\rm bound}^{(3)}-\psi^{(2\rightarrow3)}_{\rm bound}$.
As shown in Refs.~\cite{Liang2018,Chen2020}, the wave function of a three-photon weakly-bound state near $\upeta=\upzeta=0$  can be approximated by $\psi^{(3)}_{\rm bound}(\upeta,\upzeta)\propto \e^{-\sqrt{8}|\upeta|/a}\e^{-\sqrt{2}|\upeta-\sqrt{3}\upzeta|/a}\e^{-\sqrt{2}|\upeta+\sqrt{3}\upzeta|/a}$. This form can be interpreted as a product of confined states, with three pairs attracting each other simultaneously. It is then immediate to check, as shown in Fig.~\ref{fig:3}f, that the interference between the first and last terms in Eq.~\eqref{eq:simple4} produces a vortex ring with the same topology we observe experimentally.

Another important result revealed by the $\phi^{(3)}$ data in Fig.~\ref{fig:3} is that the vortex lines and ring appear approximately at the same value of $\OD\approx 80$. This observation is far from obvious, allowing us to estimate the relation between the bound state energies $E_3$ and $E_2$. An analytical model with only pairwise interaction $V^{(2)}$ predicts $E_3=4E_2$~\cite{McGuire1964}, which would result in the vortex ring appearing at a lower OD. Fitting Eq.~(\ref{eq:simple4}) to the experimental data yields  $E_3=3E_2$, so that 
 $e^{-\rmi E_3R}=-1$ forms a vortex ring  simultaneously with $e^{-\rmi E_2R}=-1$ forming the vortex lines, as shown in Fig.~\ref{fig:3}f. The result $|E_3/E_2|<4$ is an experimental evidence of a genuine three-photon repulsion term $V^{(3)}(\eta,\zeta)$ that attenuates the pairwise photon attraction~\cite{Jachymski2016,GullansPRL2016}. 
This attenuation follows from the physics of the Rydberg blockade, where one photon can `saturate' the interaction and simultaneously block two (or more) other photons.

The conditional (same-time) phase $\phi^{(3)}_0\equiv\phi^{(3)}(0,0)$, originating from the constructive sum of the vortex lines and ring, is $\phi^{(3)}_0\approx -2\pi$. This nonzero value is evident from the monotonic increase of the phase versus OD in Fig.~\ref{fig:3}b,c or from the phase winding directions in Fig.~\ref{fig:3}d. Alternatively, one can unwrap the three-dimensional $\phi^{(3)}$ data, as shown in Extended Data Fig.~\ref{fig:ED3}, to find $-2\pi$ at the center of the ring. Importantly, given that $\phi^{(2)}_0\equiv\phi^{(2)}(0)\approx -\pi$, we find $\phi^{(3)}_0=2\phi^{(2)}_0$. This result deviates from the known nonlinear quantum-Kerr phase for $n$ photons $\phi^{(n)}_\mathrm{Kerr}=\phi^{(2)}_\mathrm{Kerr} n(n-1)/2$, which ascribes $\phi^{(3)}_\mathrm{Kerr}=-3\pi$ for $\phi^{(2)}_\mathrm{Kerr}=-\pi$ \cite{haroche2006exploring}. 
Our result $\phi^{(3)}_0=2\phi^{(2)}_0$ agrees with the prediction for saturated interaction~\cite{Liang2018}, again indicating the attenuation of the interaction by the blockade saturation.

The strong three-photon attraction in the $\lambda>1$ regime, and the repulsive effect of the blockade saturation, are also evident in the three-photon correlations $g^{(3)}$ presented in Fig.~\ref{fig:4}. While the six crests of $g^{(3)}>1$ away from the center are governed by the two-photon quasibound states, the structure at the center is substantially modified by the three-photon bound state. To see this, compare the measured $g^{(3)}$ data in Fig.~\ref{fig:4}a-c to that expected from only the pairwise interactions $g^{(3)}_\mathrm{d}=g^{(2)}(t_{21})+g^{(2)}(t_{13})+g^{(2)}(t_{32})-2$ in Fig.~\ref{fig:4}d-f. $g^{(3)}_\mathrm{d}$ is the so-called disconnected part of the correlations \cite{Jachymski2016}. It is evident in all cross-sections that the central peak in $g^{(3)}$, reaching a maximal value of $g^{(3)}(0,0)=7.4\pm0.3$, is twice narrower than the corresponding peak in $g^{(3)}_\mathrm{d}$. This strong three-photon bunching has been characterized before and attributed to the tighter confinement of $\psi^{(3)}_{\rm bound}$ compared to $\psi^{(2)}_{\rm bound}$ \cite{Liang2018,Jachymski2016}. 

The effect of blockade saturation is best captured by the connected part of the correlations $g^{(3)}_\mathrm{c}=g^{(3)}-g^{(3)}_\mathrm{d}$ in Fig.~\ref{fig:4}h-i. 
We can interpret $g^{(3)}_\mathrm{c}$ as a manifestation of the contribution of the three-photon correction term $V^{(3)}$, which is repulsive. First, the repulsion attenuates the bunching of simultaneous photons, as evident by the negative $g^{(3)}_\mathrm{c}$ close to the center (blue in the figure). We observe a minimal value of $g^{(3)}_\mathrm{c,min}=-2.5\pm0.2$.
Second, as photons are less attracted to the center, we observe a ring of positive $g^{(3)}_\mathrm{c}$ (at a radius $\sqrt{\eta^2+\zeta^2}\approx 0.25~\upmu s$, red in the figure), reaching $g^{(3)}_\mathrm{c,max}=0.9\pm0.1$. As expected, with $\lambda>1$, both $g^{(3)}_\mathrm{c,min}$ and $g^{(3)}_\mathrm{c,max}$ are an order of magnitude larger than obtained previously with $\lambda\ll 1$ \cite{Liang2018,Hofferberth2018,Porto2021}.  
Finally, $g^{(3)}_\mathrm{c}$ is modulated along that ring, exhibiting six peaks where the three photons are uniformly separated in time (\textit{e.g.} along $\eta$, for $\zeta=0$). This too is due to the repulsive correction, which favors equally-spaced photons over the asymmetric arrangement of one photon separated from a close pair (\textit{e.g.} along $\zeta$, for $\eta=0$). For larger photon separations, the regularization effect of $V^{(3)}$ diminishes, and $g^{(3)}_\mathrm{c}$ approaches 0.

\section*{Discussion and summary}

We have observed topological defects, including quantum vortex-antivortex pairs, vortex lines, and vortex rings, in the few-body wavefunction of interacting photons. This is enabled predominantly by a record-strong interaction between the photons, quantified by the dimensionless parameter $\lambda$, which is the ratio between the characteristic interaction and kinetic energies. While in previous works this interaction strength was less than unity, our system reaches $\lambda\approx 3$.

The vortices in the $n$-photon wavefunction determine the maximal conditional phase $\phi^{(n)}_0$ achievable for co-propagating photons. For two photons, the vortex-antivortex pair develops when $\phi^{(2)}_0\approx-\pi$. For three photons, these pairs connect to form six vortex lines surrounding a vortex ring, the sum of which gives $\phi^{(3)}_0\approx -2\pi$. The deviation from the value $-3\pi$ expected for a quantum-Kerr nonlinearity is a manifestation of the blockade saturation. It demonstrates the role played by the vortices in controlling the conditional phase, with important consequences for deterministic quantum logic operations \cite{MolmerPRL2001,NoriPRA2010} and other quantum nonlinear devices \cite{geller2021fusing,OpatrnySciAdv2023} with three or more photons. 

Our findings are far from exhausting the potential of exploring many-body polariton physics. Even for relatively weak interactions, one could imagine exotic many-body bound states, such as a cascade of Efimov three-body states with a scaling-invariant spectrum. While initially proposed for three-dimensional structures \cite{Naidon2017}, Efimov states were later predicted to exist for a lower number of spatial dimensions \cite{Nishida2011,Petrov2015}, including quasi-one-dimensional Rydberg atom clouds \cite{Gorshkov2017}, but were not observed so far.  Our observation of the genuine three-body interaction $V^{(3)}$ could be key to the realization of four-body interactions, which are important for lattice simulations of gauge field theories~\cite{Cirac2022}.

Finally, a two-fold increase in the atomic density will realize the regime $\lambda > \pi^2$, where higher-order two- and three-photon bound states should appear \cite{BieniasPRA2014,FleischhauerPRA2017}. A superposition of vortex series arising from different bound states would drive complex nonperiodic dynamics of the few-photon wavefunction with an intricate phase structure. The three-photon vortex rings we observe are a relatively simple example of how the phase of the $n$-photon wavefunction can be controlled by an additional ($n+1$) photon. Potentially, these topological phase structures in higher dimensions could be harnessed to develop new multi-photon control tools and deterministic quantum logic.

\newpage
\begin{acknowledgments}
We thank Chen Avinadav for helpful discussions. We acknowledge financial support from the Israel Science Foundation, the US-Israel Binational Science Foundation (BSF) and US National Science Foundation (NSF), the
European Research Council starting investigator grant QPHOTONICS 678674, the Minerva Foundation with funding from the Federal German Ministry for Education and Research, the Estate of Louise Yasgour, and the Laboratory in Memory of Leon and Blacky Broder.
\end{acknowledgments}

\section*{Methods}
\noindent\textbf{Preparation of the atomic ensemble}

Our experiment starts by cooling and trapping \ce{^{87}Rb} atoms using a magneto-optical trap and then transferring them into a far-detuned optical dipole trap (ODT). The ODT consists of two 7-Watt beams ($1064~\text{nm}$) with a waist radius of $55~\upmu\text{m}$, intersecting at an angle of $37\degree$. 
A 160-MHz frequency difference between the two beams prevents spatial modulations of the ODT (here and throughout, $1~\MHz\equiv 10^6\cdot 2\pi~\mathrm{rad}/ \mathrm{s}$). The transfer of atoms from the MOT to the ODT is accompanied by further compression and cooling of the atomic cloud. To optimize these, we reduce the power of the MOT beams and increase their detuning. We find the optimization to be most sensitive to the power of the repump light. To further improve the compression, we increase the magnetic-field gradient from 15 to 22 G/cm during the transfer. We obtain $6.5\times10^5$ atoms inside the ODT, with a Gaussian density distribution of ($1\sigma$) $10\times10\times30~\upmu\text{m}^3$.

After loading the ODT, we set the quantization axis of the experiment by switching on a 5.6-G magnetic field along the optical axis (the long axis of the cloud). We then use circularly-polarized pump beam (5 nW, $F=2\rightarrow{}F'=2$) and repump beam (280 nW, $F=1\rightarrow{}F'=2$), both with a 50-$\upmu\text{m}$ waist, to optically pump the atoms to the maximal spin state ($F=2$, $m_F=2$) with $\sim 97\%$ efficiency. To avoid inhomogeneous broadening of the optical transitions inside the ODT, we periodically modulate the ODT with 10-$\upmu\text{s}$ `windows' of off-time, separated by 10-$\upmu\text{s}$ intervals of on-time. We optically pump only during the off time, over 100 such windows. 

We continue to modulate the ODT through the rest of the experiment, for 0.8 s. During the off-time windows (40000 in total), we send the probe and control light to generate Rydberg polaritons and collect data. This modulation keeps the atoms trapped (albeit with a reduced average trap depth) while avoiding the anti-trapping of excited Rydberg atoms (and other broadening mechanisms) due to the ODT. 
The decay of OD during the experiment is approximately exponential and characterized by auxiliary, $5S_{1/2}$-$5P_{3/2}$ spectroscopic measurements, performed with the control field far detuned (1.5 GHz above the resonance, thus decoupling from the Rydberg level while including the anti-trapping effect of the control field). 

\medskip
\noindent\textbf{Rydberg polaritons}

We generate polaritons using electromagnetically-induced transparency (EIT) in a ladder-type scheme, with the ground state $5S_{1/2};F=2;m_F=2$, the intermediate state $5P_{3/2};F=3;m_F=3$, and the Rydberg state $100S_{1/2};J=1/2;m_J=1/2$. 
Our probe field at 780 nm ($5S_{1/2}$-$5P_{3/2}$) originates from an external-cavity diode laser (Toptica DL-pro) and is phase-locked to a narrow-linewidth (1 kHz) frequency-doubled 1560-nm laser, which in turn is referenced to an ultra-stable high-finesse cavity (Stable Laser Systems, SLS).
The probe beam is first collimated to 10-mm diameter and set to be $\sigma^+$ polarized. It is then focused onto the cloud by a $f=60$ mm diffraction-limited aspheric lens (Thorlabs) to a 3.5-$\upmu\text{m}$ waist radius -- much smaller than $\rb$ -- such as to render the dynamics effectively one-dimensional. 

After the vacuum cell, the probe is collected by an identical aspheric lens.
It is then split into two beams, the first beam coupled to a polarization-maintaining fiber (Thorlabs 	
P3-780PM-FC-2) before being combined with the local oscillator via a 99:1 fiber combiner (PN780R1A2). The combined light is detected by a single-photon counting module (SPCM, Excelitas 780-14-FC).
The second part is coupled to a $1550$-nm polarization-maintaining fiber (P3-1550PM-FC-2) and split into three multi-mode fibers (M31L02-OM1) coupled to three SPCMs.

The control laser at 480 nm ($5P_{3/2}$-$100S_{1/2}$) is a continuous Ti-Saph laser ($\mathrm{M}^2$ lasers, SolsTiS 5000 PSX F+ECD 1000) operating at $958~\text{nm}$ and frequency doubled to $479~\text{nm}$ with an output power of 1.1 W. A small portion of the $958~\text{nm}$ light is phase modulated (using iXblue MPZ-LN-10), and a modulation sideband is locked to another ultra-stable cavity (SLS), narrowing the control laser down to a linewidth of 20 kHz. 
The control beam is integrated into the optical path using a dichroic mirror (Thorlabs DMLP550L). It is polarized before the mirror such as to drive the $\sigma^-$ atomic transition. 
The beam is carefully adjusted to focus down, through the aspheric lens, to a waist of 20 $\upmu{}$m, at the position of the probe's focus. The maximal power reaching the atoms is $450~\text{mW}$. 

We characterize the EIT parameters by fitting a ladder-type EIT model to measured transmission spectra. An example of a spectrum with high OD ($\OD=135.5\pm1.1$) is presented in Fig.~SI.2a. For the experiments in Figs.~\ref{fig:vortex}-\ref{fig:4}, we obtain the decoherence rate of the ground-to-Rydberg excitation $\gamma=70\pm 5$ kHz, the Rabi frequency of the control field $\Oc=9.5\pm0.1$ MHz, and the control field detuning $\Deltac=-27.47\pm0.03$~MHz (for Figs.~\ref{fig:vortex} and \ref{fig:3}) or $\Deltac=-28.9\pm0.03$~MHz (for Fig.~\ref{fig:4}).
Note that here and throughout the paper, decay and Rabi frequencies are in half-width convention. 
One can use the fitted parameters to calculate the accumulated phase and transmission spectra for the three-level system and the two-level system, the latter emulating the blockade situation, as shown in Fig.~SI.2b. 

The overall (two-photon) detuning from the Rydberg level is $\delta=\Delta+\Deltac$. Rydberg polaritons are formed in the vicinity of EIT, at $\delta\approx 0$. The detuning $\delta_\mathrm{TE}$ where the three-level and two-level transmission spectra cross is denoted as `transmission equality' (TE). It is given by 
\begin{equation}\label{eq: TE}
    \delta_\mathrm{TE}=\frac{(\Gamma+\gamma)\Deltac+\sqrt{\Gamma^2\Deltac^2+\Gamma(2\Gamma+\gamma)(\Gamma\gamma+\Oc^2)}}{2\Gamma+\gamma},
\end{equation}
where $\Gamma=3.03$~MHz is the half-width of the $5P$ level.
At $\delta=\delta_\mathrm{TE}$, the dissipation of blockaded and unblockaded photons is the same, and the interaction between photons becomes purely dispersive \cite{firstenberg2013attractive}. In our experiment, $\delta=0.83\delta_\textrm{TE}=1.03$~MHz (for Figs.~\ref{fig:vortex} and \ref{fig:3}, $\Delta=28.5\pm0.03$ MHz) or $\delta=1.04 \delta_\textrm{TE} =1.2$~MHz (for Fig.~\ref{fig:4}, $\Delta=27.1\pm0.03$ MHz). 
The exact detuning $\delta$ affects the interaction shift $U$ and effective mass $m$, which are both proportional to the parameter $1\le q\le 2$. As shown in SI,  
\[
\frac{1}{q}\approx{1-\frac{\delta \Delta}{\Oc^2}+\frac{2\delta}{\Delta}-\frac{4\delta^2}{\Oc^2}}\:. 
\]
We find that $q$ is maximal at the TE point, and, for our parameters, $q\approx1.4$.

For two polaritons propagating close to one another, the Rydberg energy level is shifted by $\delta E_{\text{Ryd}}=-\frac{C_6}{r^6}$, where $C_6$ is the van-der-Waals coefficient, and $r$ is the distance between the polaritons. The blockade radius $\rb=\left(\frac{q}{2}\frac{|C_6|/\hbar}{\gamma+\Oc^2/|\Delta|}\right)^{1/6}$ is the distance $r$ at which $|\delta E_{\text{Ryd}}/\hbar|$ equals the full EIT linewidth. For $r<\rb$, only one polariton can be excited. For our experimental parameters, $C_6/
\hbar=-5.617\times10^7~\MHz\cdot\upmu\text{m}^6$, and $\rb=15.3~\upmu\text{m}$. Note that we adopt the convention of definitions from Refs.~\cite{BieniasPRA2014,Jachymski2016,GullansPRL2016,Porto2021}, in which $\rb$ and $\ODb$ are $2^{1/6}$ smaller than the convention used in Refs.~\cite{Peyronel2012,firstenberg2013attractive}. 

\medskip
\noindent\textbf{Nonlinear phase measurements}

We use an interferometer to extract the nonlinear phases $\phi^{(2)}$ and $\phi^{(3)}$. One arm of the interferometer comprises the probe photons traversing the atomic medium, and the other arm functions as LO, which is frequency offset by $\nu=0.9$ MHz from the probe. The combined photon flux in the two arms is $0.3~\text{photons}/\upmu\text{s}$.

To generate the frequency offset, the probe and the LO are split from the same laser, passed through different acousto-optic modulators (AOMs), and injected into fibers.  
The RF signals driving these AOMs originate from a single multi-channel digital synthesizer. The relative phase between the arms, governed by drifts in the separate optical paths, is stable over $>20$ ms. The phase acquired by the probe in the medium in the absence of the optical nonlinearity, denoted as the linear phase $\phi^{(1)}$, depends on the OD and therefore changes during the experimental cycle. 
We extract it during each cycle in time segments of 20 ms. However, due to the limited photon rate, the extraction of $\phi^{(2)}$ and $\phi^{(3)}$ requires averaging over multiple cycles. To account for the phase drifts between cycles, we shift the time-base of each 20-ms segments by $\phi^{(1)}/\nu$. 
This shift synchronizes the beat signal in different cycles and allows to combine their data. 

For $\phi^{(2)}$, we measure the conditional beat $B^{(2)}(t_\phi,\tau)$, shown in Extended Data Fig.~\ref{fig:ED1}b, by binning two-photon detection events of one click at $t_\phi$ in the interferometer detector $D_\phi$ and one click at $t_3=\tau+t_\phi$ in detector $D_3$ (we use $D_1$ and $D_2$ as well). 
For each $\tau$ column, we fit the beat signal $B^{(2)}$ along $t_{\phi}$ to the function $a+b\cos{[\nu t_\phi+\phi^{(2)}(\tau)]}$, extracting the two-photon phase $\phi^{(2)}(\tau)$.
For $\phi^{(3)}$, we collect three-photon detection events into a three-dimensional array of time bins $B^{(3)}(t_\phi,\tau_2,\tau_3$), where $\tau_j=t_j-t_\phi$. 
For each $\{\tau_2,\tau_3\}$ column, we fit the beat signal $B^{(3)}$ along $t_{\phi}$ to the function $a+b\cos[\nu t_\phi+\phi^{(3)}_*(\tau_2,\tau_3)]$, extracting $\phi^{(3)}_*$. Finally, the three-photon phase is given by $\phi^{(3)}(\tau_2,\tau_3)=\phi^{(3)}_*(\tau_2,\tau_3)+\phi^{(2)}(\tau_3-\tau_2)$ \cite{Liang2018}.  

To produce the isosurfaces in Fig.~\ref{fig:3}d, we numerically calculate the two-dimensional gradient $\nabla\phi^{(3)}$. To do this, due to the cyclic nature of the phase angle, we first generate $P=\mathrm{exp}[\rmi\phi^{(3)}]$, smooth $P$ by a moving nearest-neighbor average to reduce noise, and then calculate $|\nabla\phi^{(3)}|=|\nabla P|$. We do this separately for each OD, in the original coordinates $\{\tau_2,\tau_3\}$, and then transform to the Jacobi coordinates $\{\eta,\zeta\}$. We choose the value $|\nabla\phi^{(3)}|=0.7~\text{rad}/\upmu s$ for the isosurface definition, generating an isosurface structure that is large enough to show detail around the vortex but small enough to delineate the vortex paths faithfully. The colormap projected on the surfaces is $\phi^{(3)}$ at their positions. 

\bibliography{bibliography}

\renewcommand{\figurename}{\textbf{Extended Data Fig.}}

\setcounter{figure}{0}
\renewcommand{\thefigure}{\textbf{\arabic{figure}}}
\renewcommand{\theHfigure}{E\arabic{figure}}

\begin{figure}[h]
\centering\includegraphics[width=0.75\columnwidth,trim={0cm 3.5cm 17cm 3.5cm},clip]{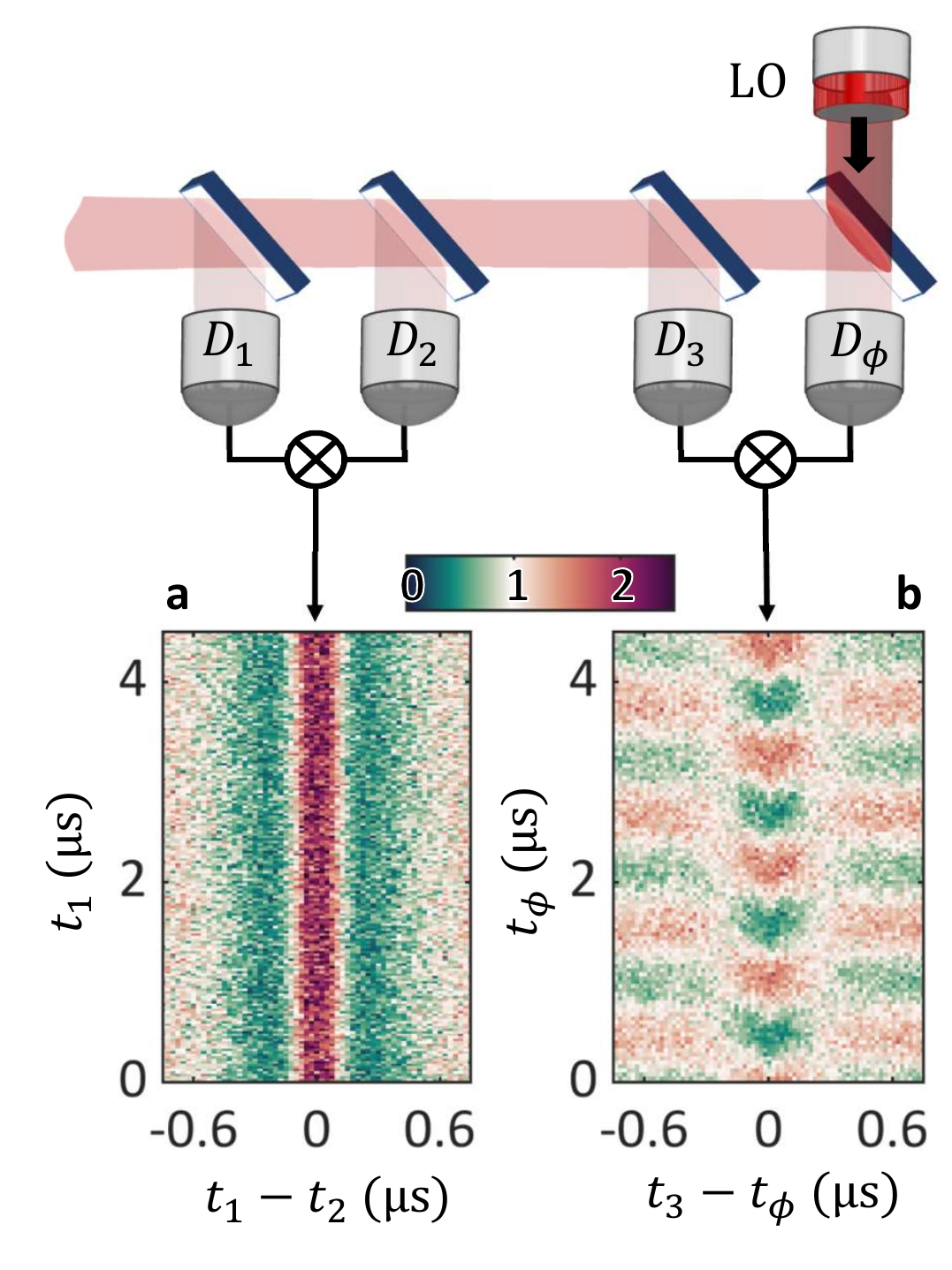}
\caption{{\bf Example for raw data, used for extracting the two-photon normalized correlation function $g^{(2)}(\tau)$ and conditional phase $\phi^{(2)}(\tau)$.}
After traversing the medium, the probe light is split and measured by the single-photon detectors $D_{1-3,\phi}$. We denote the respective detection times by $t_{1-3,\phi}$ \textbf{(a)} Normalized correlations $g^{(2)}(t_1,\tau=t_1-t_2)$, extracted from detectors $D_{1,2}$. Averaging over $t_1$ then provides $g^{(2)}(\tau)$. In practice, we use the three pair-combinations of $D_{1-3}$. For the three-photon correlation function $g^{(3)}$, we use all $D_{1-3}$. \textbf{(b)} The beat signal $B^{(2)}(t_\phi,\tau=t_3-t_\phi)$. Detector $D_\phi$ measures the interference beat between the probe and a frequency-shifted local oscillator (LO). Conditioning the beat detections on detections in $D_{3}$ (and also in $D_{1,2}$) provides $B^{(2)}$. The phase of the beat signal for each $\tau$ is $\phi^{(2)}(\tau)$. In the example shown, $\phi^{(2)}(0)=\pi$. For the three-photon conditional-phase $\phi^{(3)}$, we use $D_\phi$ and the three pair-combinations of $D_{1-3}$.
}\label{fig:ED1}
\end{figure}

\begin{figure}[H]
\centering\includegraphics[width=\columnwidth, trim={0cm 0 1cm 0cm},clip]{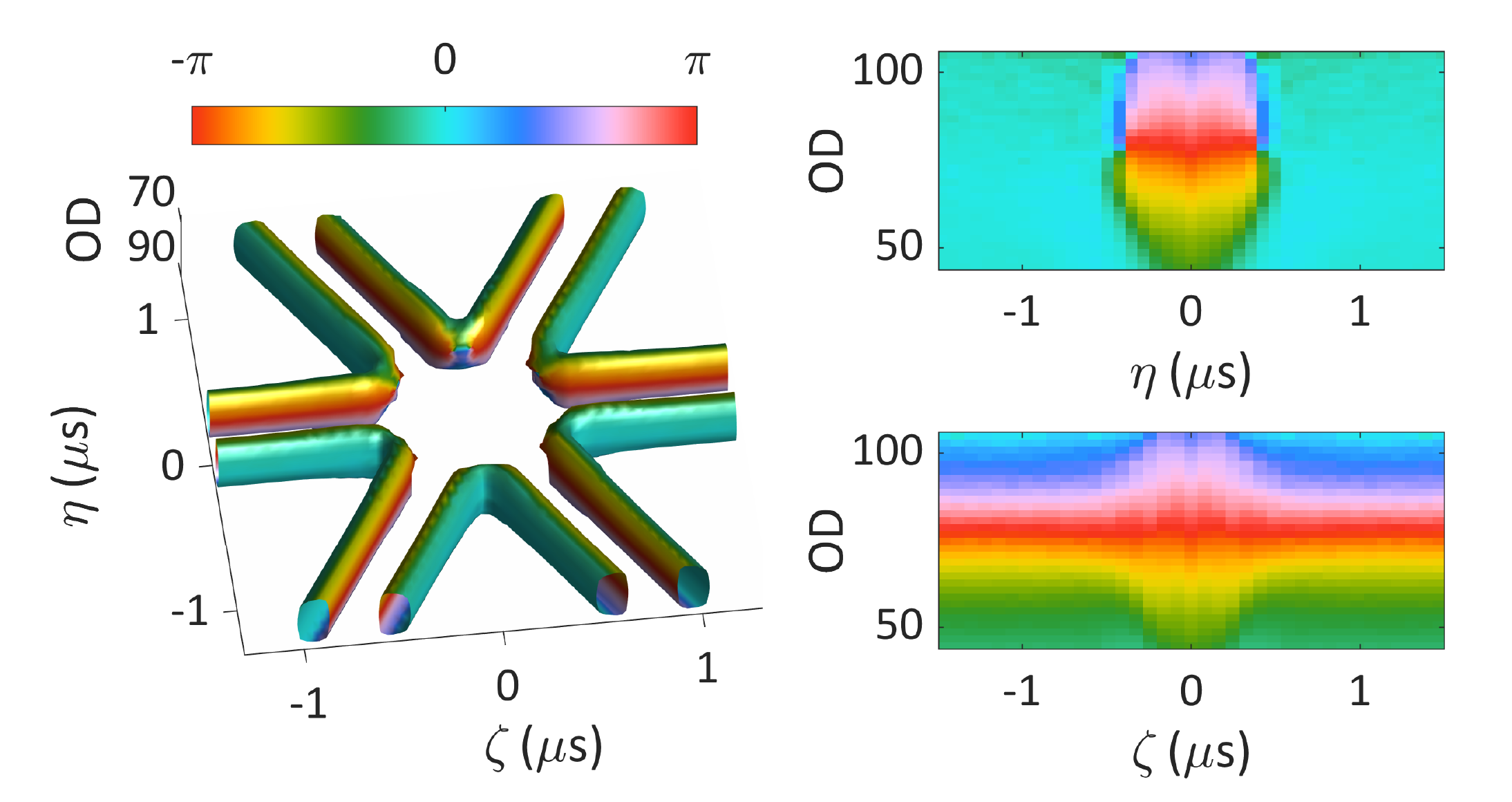}
\caption{\textbf{Expected three-photon phase in the absence of three-photon interactions}, $\arg[\Psi^{(2\rightarrow3)}]$, extrapolated from the measured two-photon correlations $g^{(2)}(\tau)$ and phase $\phi^{(2)}(\tau)$. Inspired by $\psi^{(2\rightarrow3)}_{\rm bound}$, we use the ansatz
$\Psi^{(2\rightarrow3)}=\Psi^{(2)}(t_{21})+\Psi^{(2)}(t_{13})+\Psi^{(2)}(t_{32})-2$, where $\Psi^{(2)}(\tau)=\sqrt{g^{(2)}(\tau)}\e^{\rmi \phi^{(2)}(\tau)}$.
The constant term $-2$ ensures that the photons are uncorrelated, $|\Psi^{(2\rightarrow3)}|=1$, when all three times $t_{1,2,3}$ strongly differ. 
We find a star-like structure of vortex lines, similar to that in the actual three-photon data, while the vortex ring, associated with the three-photon bound state, is absent (compare to Fig.~\ref{fig:3}b-d).}
\label{fig:ED2}
\end{figure}

\begin{figure}[H]
\centering\includegraphics[width=0.5\columnwidth, trim={0cm 0cm 0cm 0cm},clip]{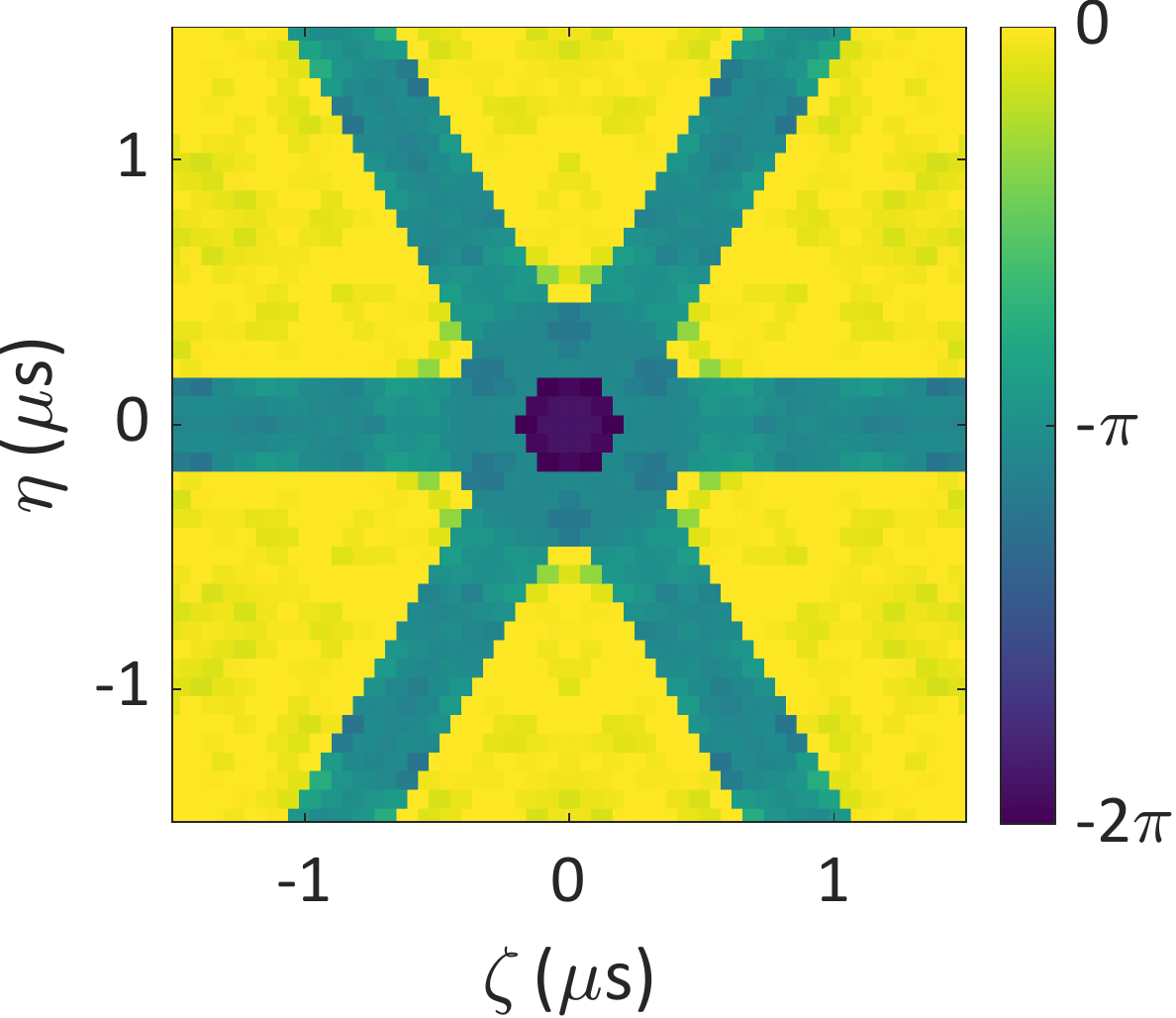}
\caption{\textbf{Unwraped three-photon phase $\phi^{(3)}(\eta,\zeta)$} (same cross-section as in Fig.~\ref{fig:3}a). While two co-propagating photons acquire a $\pi$ conditional phase, three co-propagating photons acquire $2\pi$.
}
\label{fig:ED3}
\end{figure}

\end{document}


\title{Supplementary Information for \\ ``Quantum vortices of strongly interacting photons''}

\author{Lee Drori}
\author{Bankim Chandra Das}
\author{Tomer Danino Zohar}
\author{Gal Winer}
\author{\linebreak Eilon Poem}
\author{Alexander Poddubny}
\author{Ofer Firstenberg}
\affiliation{Department of Physics of Complex Systems, Weizmann Institute of Science, Rehovot 7610001, Israel}

\maketitle
\renewcommand{\thefigure}{SI.\arabic{figure}}

\section{Numerical and analytical theory}
\subsection{Physical model and numerical approach}\label{sec:numerics}
The numerical methods used to simulate the experimental system follow the framework of Ref.~\cite{firstenberg2013attractive}. As our probe beam waist $w_0$ is much smaller than the Rydberg blockade radius $\rb$, we consider a one-dimensional system of three levels, with the creation operators $\hat{\varepsilon}^\dagger(x)$, $\hat{p}^\dagger(x)$, and $\hat{s}^\dagger(x)$ for a photon, an intermediate-level excitation, and a Rydberg-level excitation, respectively. We assume that the operators follow the Heisenberg equations of motion and obey bosonic commutation relations, \textit {i.e.}, $\left[\hat{\varepsilon}(x_1),\hat{\varepsilon}^\dagger(x_2)\right]=\left[\hat{p}(x_1),\hat{p}(x_2)^\dagger\right]=\left[\hat{s}(x_1),\hat{s}(x_2)^\dagger\right]=\delta(x_1-x_2)$, and we will ignore the Langevin noise terms. As we are sending a weak coherent state, the wavefunction has the form
\begin{equation}\label{eq:psi}
\begin{array}{lll}
    \ket{\psi(t)}&=& \epsilon \ket{0}+ \int\limits_{-\infty}^{+\infty}dx_1 E(x_1,t)\hat{\varepsilon}^\dagger(x_1)\ket{0}+\int\limits_{-\infty}^{+\infty} dx_1 P(x_1,t)\hat{p}^\dagger(x_1)\ket{0}+\int\limits_{-\infty}^{+\infty}dx_1 S(x_1,t)\hat{s}^\dagger(x_1)\ket{0}\\
    &&
+\dfrac{1}{2}\iint\limits_{-\infty}^{+\infty}dx_1 dx_2 EE(x_1,x_2)\hat{\varepsilon}^\dagger(x_1)\hat{\varepsilon}^\dagger(x_2)\ket{0}+\dfrac{1}{2}\iint\limits_{-\infty}^{+\infty}dx_1 dx_2 \left[EP(x_1,x_2)+PE(x_2,x_1)\right]\hat{\varepsilon}^\dagger(x_1)\hat{p}^\dagger(x_2)\ket{0} \\
&&+\dfrac{1}{2}\iint\limits_{-\infty}^{+\infty}dx_1 dx_2 PP(x_1,x_2)\hat{p}^\dagger(x_1)\hat{p}^\dagger(x_2)\ket{0} +\dfrac{1}{2}\iint\limits_{-\infty}^{+\infty}dx_1 dx_2 \left[PS(x_1,x_2)+SP(x_2,x_1)\right]\hat{p}^\dagger(x_1)\hat{s}^\dagger(x_2)\ket{0}\\
&&+\dfrac{1}{2}\iint\limits_{-\infty}^{+\infty}dx_1 dx_2 SS(x_1,x_2)\hat{s}^\dagger(x_1)\hat{s}^\dagger(x_2)\ket{0}+\dfrac{1}{2}\iint\limits_{-\infty}^{+\infty}dx_1 dx_2 \left[ES(x_1,x_2)+SE(x_2,x_1)\right]\hat{\varepsilon}^\dagger(x_1)\hat{s}^\dagger(x_2)\ket{0},
 \end{array}
\end{equation}
where we assume $|\epsilon|\ll 1$ and neglect higher-order terms (of three or more excitations) and $x_1$ and $x_2$ are the coordinates of two photons.
Using the Heisenberg equation of motion for the wave function Eq.~\eqref{eq:psi}, the equations for the single-photon amplitudes become \cite{firstenberg2013attractive}
\begin{equation}\label{eq:one-photon}
\setstretch{2.25}
	\begin{array}{ll}
	\dfrac{\partial E(x)}{\partial  t} &= -c \dfrac{\partial E(x)}{dx}+ i\sqrt{\rho(x)}g P(x)\:,\\
	\dfrac{\partial P(x)}{\partial t} &= - (\Gamma- i \DeltaP)P(x)+ i  \sqrt{\rho(x)}g E(x)+ i \Oc S(x)\:,\\
	\dfrac{\partial S(x)}{\partial t} &= - (\gamma- i \delta)S(x)+ i  \Oc P(x)\:.
	\end{array}
	\end{equation}
Here $g$ is the single-photon single-atom Rabi frequency of the probe field, connected to OD via the relation $\OD=(2 g^2) / (\Gamma c) \int \rho(x) dx=(2\sqrt{2\pi}g^2\rho_0\sigma) / (\Gamma c)$, where 
$\rho(x)=\rho_0 e^{-x^2/(2\sigma^2)}$ is the Gaussian density profile of the medium along the photon propagation direction, and $c$ is the speed of light in vacuum. 

The solution for the single-photon amplitudes in the steady state reads
\begin{align}\label{steady state solution one photon}
E(x) &= \exp\biggl[-\dfrac{g^2}{\Gamma c}\dfrac{\Gamma}{\Gamma- i \DeltaP+ \dfrac{\Oc^2}{\gamma-i\delta}}\int_{-\infty}^{x}\rho(x')dx'\biggr]\:,\\
P(x)&= \dfrac{i \sqrt{\rho(x)} g }{\Gamma- i \DeltaP+ \dfrac{\Oc^2}{\gamma-i\delta}}E(x)\:,\nonumber
\\
S(x)&= -\dfrac{ \Oc \sqrt{\rho(x)} g }{(\gamma- i\delta)(\Gamma- i\DeltaP)+\Oc^2} E(x)\:.\nonumber
\end{align}
\newpage
For the two-photon amplitudes, we write a system of equations similar 
 to Eqs.~\eqref{eq:one-photon},  
\begin{equation}
\begin{array}{lll}
\dfrac{\partial EE(x_1,x_2)}{\partial t}&=& -c \left(\dfrac{\partial}{\partial x_1}+\dfrac{\partial}{\partial x_2}\right)EE(x_1,x_2) + i g \left[\sqrt{\rho(x_1)}PE(x_1,x_2)+\sqrt{\rho(x_2)}EP(x_1,x_2)\right]\:,\\
& &\\
\dfrac{\partial EP(x_1,x_2)}{\partial t}&=&- c \dfrac{\partial}{\partial x_1}EP(x_1,x_2)-(\Gamma-i\DeltaP)EP(x_1,x_2) \\
&&+ ig \left[\sqrt{\rho(x_1)}PP(x_1,x_2)+\sqrt{\rho(x_2)}EE(x_1,x_2)\right]+ i \Oc ES(x_1,x_2)\:,\\
& &\\
\dfrac{\partial ES(x_1,x_2)}{\partial t}&=&-c \dfrac{\partial}{\partial x_1}ES(x_1,x_2)-(\gamma-i\delta)ES(x_1,x_2) + i g\sqrt{\rho(x_1)}PS(x_1,x_2)\\&+& i \Oc EP(x_1,x_2)\:,\\
&&\\
\dfrac{\partial PE(x_1,x_2)}{\partial t}&=&- c \dfrac{\partial}{\partial x_2}PE(x_1,x_2)-(\Gamma-i\DeltaP)PE(x_1,x_2)\:\\ 
&&+ ig\left[\sqrt{\rho(x_2)}PP(x_1,x_2)+\sqrt{\rho(x_1)}EE(x_1,x_2)\right]+ i \Oc SE(x_1,x_2)\:,\\
&&\\
\dfrac{\partial PP(x_1,x_2)}{\partial t}&=&-2(\Gamma-i\DeltaP)PP(x_1,x_2) + i g\left[\sqrt{\rho(x_1)}EP(x_1,x_2)+\sqrt{\rho(x_2)}PE(x_1,x_2)\right]\\
&&+ i \Oc\left[PS(x_1,x_2)+ SP(x_1,x_2)\right]\:,\\

& &\\
\dfrac{\partial PS(x_1,x_2)}{\partial t}&=&-(\Gamma-i\DeltaP)PS(x_1,x_2)-(\gamma-i\delta)PS(x_1,x_2) + i g \sqrt{\rho(x_1)}ES(x_1,x_2)\\&+&i \Oc\left[SS(x_1,x_2)+PP(x_1,x_2)\right]\:,\\
& &\\

\dfrac{\partial SE(x_1,x_2)}{\partial t}&=&- c\dfrac{\partial}{\partial x_2}SE(x_1,x_2)-(\gamma-i\delta)SE(x_1,x_2) + i\Gamma\sqrt{\rho(x_2)}SP(x_1,x_2)\\&+& i \Oc PE(x_1,x_2)\:,\\
&&\\
\dfrac{\partial SP(x_1,x_2)}{\partial t}&=&-(\Gamma-i\DeltaP)SP(x_1,x_2)-(\gamma-i\delta)SP(x_1,x_2) + i g\sqrt{\rho(x_2)}SE(x_1,x_2)\\&+&i \Oc\left[SS(x_1,x_2)+PP(x_1,x_2)\right]\:,\\
&&\\
\dfrac{\partial SS(x_1,x_2)}{\partial t}&=&-2(\gamma-i\delta)SS(x_1,x_2) +i \Oc\left[PS(x_1,x_2)+SP(x_1,x_2)\right]\\&-&i V_{ss}(x_1-x_2)SS(x_1,x_2)\:,
\end{array}\label{eqs:full2}
\end{equation}
where we have added the van der Waals interaction $V_{ss}(r) = -(C_6/\hbar)/|r|^6$ in the last equation.

\begin{figure}[h!]
\centering
\includegraphics[scale=.55, trim={0cm 0cm 15cm 0cm},clip]{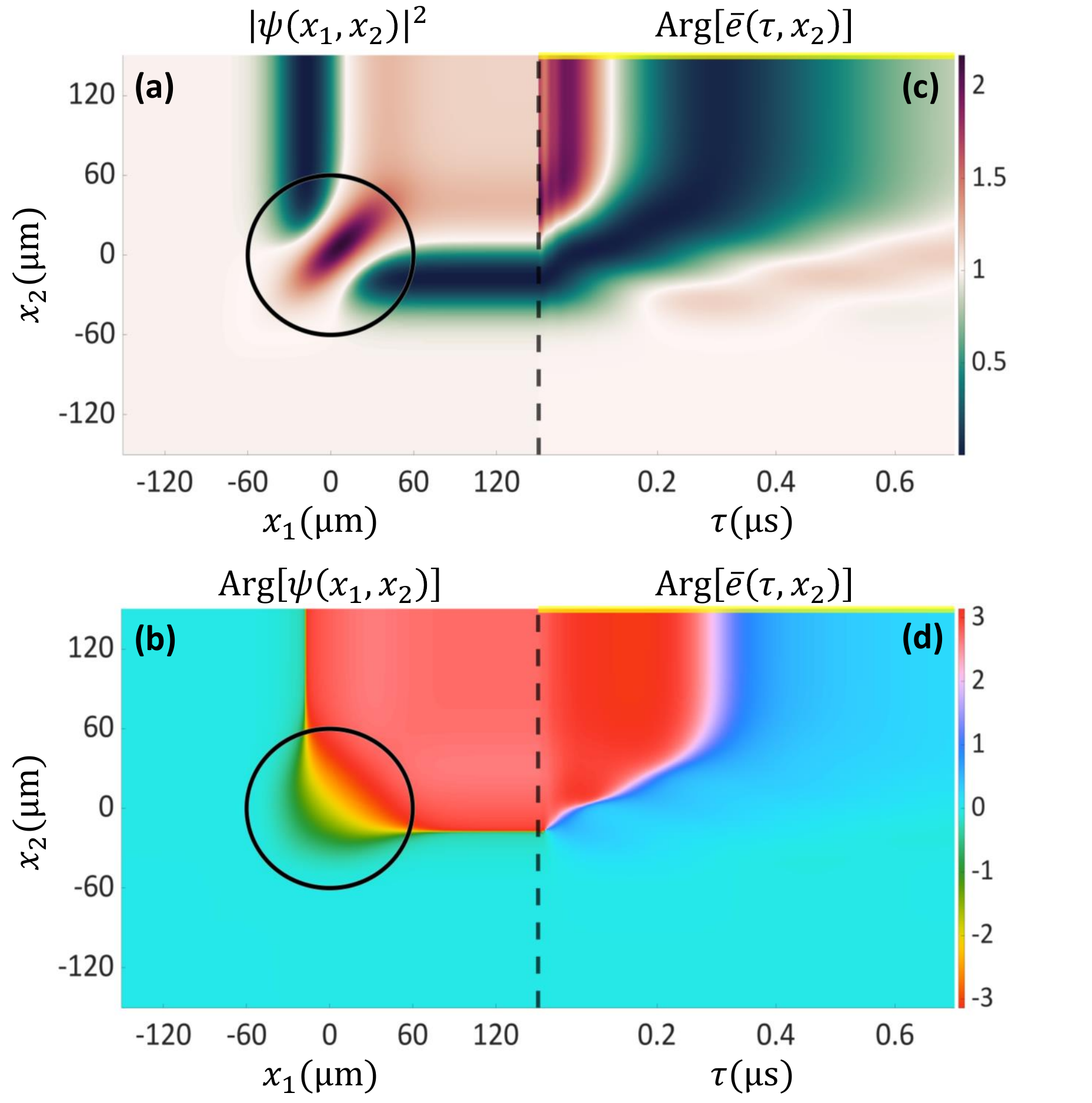}
\caption{Numerical simulations for \textbf{(a,b)} the propagation of two photons and \textbf{(c,d)} the propagation of the second photon after the detection of the first photon. \textbf{(a)} The normalized probability $|\psi(x_1,x_2)|^2$, where $\psi(x_1,x_2)=EE(x_1,x_2)/[E(x_1)E(x_2)]$. \textbf{(b)} The two-photon phase Arg$[\psi(x_1,x_2)]$. \textbf{(c)} The normalized probability $|\overline{e}(\tau,x_2)|^2$, where $\overline{e}(\tau,x_2)=e(\tau,x_2)/[E(\xo)E(x_2)]$. Here $\tau$ is the time passed from the detection of the first photon, and $\xo=5\sigma$ is well outside the cloud. \textbf{(d)} The phase Arg$[\overline{e}(\tau,x_2)]$. The values along the horizontal yellow lines correspond to the experimental observables $g^{(2)}(\tau)=|\overline{e}(\tau,\xo)|^2$ and $\phi^{(2)}(\tau)=\mathrm{Arg}[\overline{e}(\tau,\xo)]$.
 Calculation parameters are $\OD = 73$, $\sqrt{2\pi}\sigma=75~\upmu$m, $\rb=15.3~\upmu$m, $\Oc=9.5$~MHz, $\Gamma= 3.03$~MHz, $\gamma=0.07$~MHz, $\DeltaP=28.5$~MHz, and $\delta=1.03$~MHz (0.2 MHz from the TE point). The circles represent the edge ($2\sigma$) of the Gaussian cloud.
}
\label{full_numerical_gaussian}
\end{figure}

These nine equations can be readily solved numerically to get the steady-state two-photon amplitude $EE(x_1,x_2)$ within the range $-\xo<x_1<\xo$, $-\xo<x_2<\xo$, as shown in Figs.~\ref{full_numerical_gaussian}a,b. We choose $\xo=5\sigma $ to be well outside the medium and, assuming that the photons are initially uncorrelated, take the steady-state boundary conditions $EE(x_1,-\xo)=E(x_1)$ and $EE(-\xo,x_2)=E(x_2)$  from Eqs.~(\ref{steady state solution one photon}).

To calculate the experimental observables $g^{(2)}(\tau)$ and $\phi^{(2)}(\tau)$, we assume that the first photon is detected at $x_1=\xo$ at time $t=0$ and calculate the dynamics of the second photon while it propagates outside (to $x_2=\xo$) to be detected at $t=\tau$. We denote the amplitudes (of the second photon) in this calculation as $e(\tau,x_2)$, $p(\tau,x_2)$, and $s(\tau,x_2)$ and solve the equations of motion in Eq.~(\ref{eq:one-photon}) from $t=0$ to $t= \tau$, subject to the initial and boundary conditions
\begin{equation}
\begin{array}{lll}
e(0,x_2)&=& EE(x_1=\xo,x_2)\\
p(0,x_2)&=& EP(x_1=\xo,x_2)\\
s(0,x_2)&=& ES(x_1=\xo,x_2)\\
e(t,x_2=-\xo)&=& EE(x_1=\xo,x_2=-\xo)\:.
\end{array}
\end{equation}
The solution after proper normalization is shown in Figs.~\ref{full_numerical_gaussian}c,d, with the yellow lines corresponding to $g^{(2)}(\tau)$ and $\phi^{(2)}(\tau)$, where $\tau$ is the time between the two detections. The calculation shown in Fig.~\ref{full_numerical_gaussian} is done for $\OD=73$; repeating this calculation for a range of $\OD$ provides the results shown in Figs.~2c-e in the main text.

\subsection{Derivation of the Schr\"odinger equation}
The system of equations Eqs.~\eqref{eqs:full2} for the two-photon amplitudes can be readily solved numerically. It quantitatively reproduces all experimental results for two photons. However, it is too cumbersome to allow for a transparent analytical description and interpretation of the experiment.  Hence, it is instructive to obtain  approximate closed-form equations for the two-photon amplitude. Such derivation of an effective two-photon Schr\"odinger equation is presented in this section.  It generalizes the results of Refs.~\cite{firstenberg2013attractive,Peyronel2012}  for the case of arbitrary detuning $\delta\ne 0$ from the EIT resonance. Moreover, we derive a Dirac-like equation for the two-photon amplitudes that is a better approximation than the Schr\"odinger equation for a spatially inhomogeneous density profile.

We start by reintroducing the center-of-mass and relative coordinates $R= (x_1+x_2)/2$ and $r=x_1-x_2$ and define the symmetric and anti-symmetric combinations of the amplitudes
\begin{equation} ES_{\pm}= \dfrac{ES(x_1,x_2)\pm SE(x_1,x_2)}{2},\quad PS_{\pm}= \dfrac{PS(x_1,x_2)\pm PS(x_1,x_2)}{2},\quad EP_{\pm}= \dfrac{EP(x_1,x_2)\pm PE(x_1,x_2)}{2}\:.\end{equation} 
Next, we eliminate $PP$, $PS_{\pm }$, $EP_{\pm }$, and $SS$ from Eqs.~\eqref{eqs:full2}.
For simplicity we assume here that $\gamma=0$, $\Oc\ll \DeltaP$, and $\Gamma\ll \Delta$ and first consider a spatially homogeneous cloud of length $L= \sqrt{2\pi}\sigma$.
The elimination procedure results in the following Dirac-like  equation for the amplitudes $ES_{\pm}$ in the lowest order in $\Gamma/\DeltaP$:
 \begin{equation}\label{eq:dirac}
i\dfrac{\partial}{\partial R} 
\begin{pmatrix}
ES_+\\ES_-
\end{pmatrix}=
\begin{pmatrix} -2m_0(1-q_0)+V^{(2)}(r)& -\rmi \frac{\partial}{\partial r}\\
 -\rmi \frac{\partial}{\partial r} &
-2m_0
\end{pmatrix}
\begin{pmatrix}
ES_+\\ES_-
\end{pmatrix}\:.
\end{equation}
Here, the effective mass and polariton-polariton interaction potential are, respectively, \begin{equation}\label{eq:mass2}
m\equiv q_0m_0=-\frac{\OD}{8L}\frac{\Gamma  q_0}{\Delta},\quad 
 V^{(2)}(r)=\frac{\OD}{L}\frac{\Gamma q_0}{\Delta}\frac{r_{\rm b}^6}{
 r_{\rm b}^6+r^6
 },
\end{equation}
where
\vspace{-3mm}
\begin{equation}\label{eq:q}
    q_0=\frac{\Omega^2}{\Omega^2-\delta \Delta},
\end{equation}
and {$\rb=\left[q_0|C_6\Delta|/(2\hbar \Omega^2)\right]^{1/6}$} is the Rydberg blockade radius for $\gamma=0$. To obtain the Schr\"odinger equation used in the main text, we shall adiabatically eliminate $ES_-$ from Eqs.~\eqref{eq:dirac}. To this end, we introduce the new variables $\psi,\varrho$ by defining  $ES_+=\psi \e^{\rmi 2m_0(1-q_0) R}$,
$ES_-=\varrho \e^{\rmi 2m_0(1-q_0) R}$. Equations~\eqref{eq:dirac} then become 
\begin{equation}\label{eq:dirac2}
i\dfrac{\partial}{\partial R} 
\begin{pmatrix}
\psi\\\varrho
\end{pmatrix}=
\begin{pmatrix} V^{(2)}(r)& -\rmi \frac{\partial}{\partial r}\\
 -\rmi \frac{\partial}{\partial r} &
-2m
\end{pmatrix}
\begin{pmatrix}
\psi\\ \varrho
\end{pmatrix}\:.
\end{equation}
Expressing $\varrho$ from the second equation as 
$\varrho=-\frac{\rmi }{2m}\frac{\partial}{\partial r}\psi$, we obtain the form of Eq.~(1) in the main text:
\begin{equation}\label{eq:Schrodinger}
	    \rmi \dfrac{\partial}{\partial R } \psi(r,R) = -\dfrac{1}{2m}\dfrac{\partial^2}{\partial r^2}\psi(r,R) + V^{(2)}(r)\psi(r,R)\:.
	\end{equation}
The $EE(r,R)$ amplitude will follow the same equation of motion with a scale factor of $g/\Oc$~\cite{firstenberg2013attractive}. The equation is to be solved with the initial condition $\psi(r,R=0)=1$ for $R=0\ldots L$. The Schr\"odinger equation solution agrees qualitatively with the experimental results in the approximation of a homogeneous cloud.  

An inhomogeneous cloud  can be considered by replacing the $1/L$ factor in  Eqs.~\eqref{eq:mass2} with the Gaussian density profile $\e^{-R^2/(2\sigma^2)}/(\sqrt{2\pi}\sigma)$. The effective mass will then depend on $R$ and  diverge at the  cloud edges, making the Schr\"odinger equation  inapplicable. On the other hand, the Dirac equation [Eq.~\eqref{eq:dirac2}] still works well for an inhomogeneous cloud and approximates well the solution of the full system [Eqs.~\eqref{eqs:full2}].

\vspace{-2mm}
\subsection{Estimation of the interaction strength}

\vspace{-2mm}
The effective strength of the potential $\lambda$ is given by the product of the mass, potential depth, and interaction range,
\begin{equation}
    \lambda=2|m| r_{\rm b}^2 V^{(2)}(r=0)=
    \left(\frac{q_0\Gamma}{2\DeltaP}\ODb \right)^2,\quad \mathrm{where}~\ODb=\OD\frac{\rb}{L}\:.
\end{equation}
The dimensionless parameter $q_0$ in Eq.~\eqref{eq:q} strongly depends on the detuning $\delta$. At the EIT point, with $\delta=0$, we find $q_0=1$, while at the transmission equality (TE) point, with $\delta=\delta_\mathrm{TE}=\Omega^2/(2\Delta)$, we find $q_0=2$. This means that, at TE, the mass and the potential depth are both larger by a factor of two, and the interaction strength $\lambda$ is four times larger than on EIT. 

For our experimental parameters ($\Omega = 9.5~{\rm MHz}$, $\Delta = 28.5~{\rm MHz}$, $\delta=1.03~{\rm MHz}$), Eq.~\eqref{eq:q} yields the value of $q_0=1.5$. However, the derivation of Eq.~\eqref{eq:q} assumes that $\gamma=0$ and $\Gamma\ll \Omega\ll \Delta$. 
In the experiment, however, $\gamma = 0.07~{\rm MHz}$ and $\Gamma = 3.03~{\rm MHz}$. The ratios $\Delta/\Omega\approx 3$ and $\Omega/\Gamma\approx 3$ are also finite.
A more accurate numerical calculation with these parameters suggests a slightly lower value of $q=1.4$. In order to resolve this  discrepancy, it is possible to use a more precise analytical expression for the polariton mass, derived from the single-photon susceptibility $\chi$,
\begin{equation}\label{eq:m1}
m=\frac{1}{4}\frac{\left(\frac{\rmd \Re\chi}{\rmd \delta}\right)^2}{\frac{\rmd^2 \Re\chi}{\rmd \delta^2}}
\text{ with } \chi\propto\frac{\Gamma}{-\Delta_\mathrm{c}+\delta+\rmi \Gamma+\frac{\rmi \Omega^2}{\gamma-\rmi\delta}}\:,
\end{equation}
where $\Delta_\mathrm{c}=\delta-\Delta$ is the control-field detuning.
The inverse mass is proportional to the curvature of the polariton dispersion, and the numerator in Eq.~\eqref{eq:m1}, depending on  the polariton group velocity, is the normalization factor.
For $\Gamma,\gamma\ll \Omega\ll \Delta$, using Eq.~\eqref{eq:m1}, we arrive at the enhancement factor
\begin{equation}\label{eq:q2}
q\equiv \frac{m(\delta)}{m(\delta=0)}\approx\frac1{1-\frac{\delta \Delta}{\Omega^2}+\frac{2\delta}{\Delta}-\frac{4\delta^2}{\Omega^2}}\:.
\end{equation}
This expression results in the value of 
$q=1.4$ at the work point, in agreement with the numerics.

\subsection{Weak interaction limit}\label{sec:weak}
 In order to obtain a further understanding of the vortex dynamics, it is instructive to consider separately  the regime of weak interaction, $\lambda \ll 1$. In this case, the actual potential can be well approximated by a $\delta$-shaped potential well. Instead of Eq.~\eqref{eq:Schrodinger}, we write
\begin{equation}\label{eq:Schrodinger1}
	    i \dfrac{\partial}{\partial R } \psi(r,R) = -\dfrac{1}{2m}\dfrac{\partial^2}{\partial r^2}\psi(r,R) + \frac{\lambda}{m r_{\rm b}} \delta(r)\psi(r,R)\:.
	\end{equation}
 This equation is to be solved with the homogeneous initial condition $\psi(r,R=0)=1$.
Following Ref.~\cite{firstenberg2013attractive}
we expand the dynamics as a sum over the bound stationary eigenstate $\psi_{\rm bound}(r)$ with the energy 
$ E_{2}=-\lambda^2/(2mr_{\rm b}^2)$
and the continuous spectrum eigenstates $\{\psi_k,\varepsilon_k\}$
\begin{equation}
\psi(r,R)=c \e^{-\rmi E_{2} R} \psi_{\rm bound}(r)+\int_{0}^{\infty} \rmd k c_k \e^{-\rmi \varepsilon_k R} \psi_k (r)
\end{equation}
with the energies $\varepsilon_k=k^2/(2m)$.
The result is given by \cite{firstenberg2013attractive}
\begin{equation}\label{eq:expansion}
\psi(r,R)=2\e^{-\rmi E_{2}R-|z|}+\frac{2}{\pi}\int\limits _0^\infty \rmd k\frac{\sin (kz)-(k r_{\rm b}/\lambda)\cos (kz)}{ k[(k r_{\rm b}/\lambda)^2+1]}\e^{-\rmi \varepsilon_kR},\quad z=\frac{\lambda |r|}{ r_{\rm b}}\:.
\end{equation}
The expression under the integral quickly decays at large wave vectors $k$. This reflects the weak overlap of the homogeneous initial wavefunction with the large-$k$ scattering states. Thus, the continuous-spectrum term is determined by the  extended, small-$k$ scattering states with low energies $\varepsilon_k$. Hence, the phase of the continuous-spectrum term $\propto \varepsilon_k R$ changes only slowly with $R$. On the other hand, the phase of the bound-state term increases with $R$ faster, as imposed by the bound-state energy $E_2$.  In the crudest approximation, the phase accumulation of the scattering states can be entirely neglected, which results in Eq.~(2) in the main text. More accurate consideration is presented below.

We are interested in relatively large values of $|r|$ and $R$, for which the integral in Eq.~(\ref{eq:expansion}) is dominated by small $k$, and the term
$(k r_{\rm b}/\lambda)^2$ in the denominator can be neglected.  The integral can then be evaluated analytically with the help of the   special function
\newcommand{\erfi}{\mathop\mathrm{erfi}}
$\erfi(y)\equiv -\rmi \mathop\mathrm{erf}(\rmi y)$,
\begin{equation}\label{eq:erfi}
    \psi(r,R)=2\e^{-\rmi E_{2}R-z}+\rmi \,{\rm erfi} \sqrt{\frac{-\rmi z^2}{4  E_{2}R}}-\sqrt{\frac{\rmi }{{\pi E_{2}R}}}\e^{-\frac{\rmi z^2}{4  E_{2}R}}\:.
\end{equation}
Since the  vortices form where the phase $E_{2}R$ is large, we are interested in the regime where the argument of the $\erfi$ function in Eq.~\eqref{eq:erfi} is small. Using the Taylor expansion $\erfi y\approx 2y/\sqrt{\pi}$, we simplify Eq.~\eqref{eq:erfi} to
\begin{equation}\label{eq:an}
\psi\approx 2\e^{-\rmi E_{2}R-\lambda|r|/r_{\rm b}}-\sqrt\frac{{\rmi}}{{\pi  E_{2}R}}\left(1-\frac{\lambda |r|}{ r_{\rm b}}\right)\:.
\end{equation}
Here the first term corresponds to the bound state and the second one is the contribution of the continuous spectrum. As compared to the simpler approximation, Eq.~(2) in the main text, the continuous spectrum contribution in Eq.~\eqref{eq:an} has an extra nonzero phase of $3\pi/4$ at $r=0$. Similarly to Eq.~(2), this phase does not depend on $R$. 
Equation~\eqref{eq:an} has pairs of vortices and antivortices for $R=R_k$, where
 \begin{equation}\label{eq:positions1}
 E_2R_k=3\pi/4+2\pi k, ~~ k=0,1,2\ldots,
 \end{equation}
 and 
 \begin{equation}\label{eq:r0}
 \frac{\lambda|r|}{r_b}\equiv z_0=  W\left(\frac{\pi\sqrt{3}}{\e}\right)+1\approx 1.853\:.
\end{equation} 
where $W$ is the Lambert W-function defined by the transcendental equation $W(y)\e^{W(y)}=y$.
In order to demonstrate this, we expand Eq.~\eqref{eq:an} up to linear terms in $R$ and $|r|$
near the points $R=R_k$, $\lambda |r|/r_b=z_0$:
\begin{equation}\label{eq:an1}
\psi(R,r)\approx \delta R E_2(R-R_k)+\delta z \left(\frac{\lambda|r|}{r_b}-z_0\right)\:,
\end{equation}
where 
\begin{equation}\label{eq:deltaz}
\delta z=\frac{2z_0}{\pi}\sqrt{\frac{\rmi}{3}}\approx 0.48+0.48\rmi,\quad \delta R=
\frac{2(z_0-1)}{\pi^2}\sqrt{\frac{\rmi}{3}}\left(\rmi \pi-\frac{2}{3}\right)\approx -0.27+0.17\rmi \:.
\end{equation}
The coefficients $\delta R$ and $\delta z$ in Eq.~\eqref{eq:an1} have different phases, which means that Eq.~\eqref{eq:an1} describes a phase vortex in the $(R,r)$ space.

While Eq.~\eqref{eq:an} has been derived for the $\delta$-potential approximation, which is applicable only to weakly bound states, $\lambda\ll 1$,  it can also be used to find the values of $R$ where vortices occur even when $\lambda\gtrsim 1$.  For  $\lambda \gtrsim 1$, the expression $E_2=-\lambda/(2m\rb^2)$ for the bound state energy in Eq.~\eqref{eq:an} is no longer valid. Instead, one should substitute into Eq.~\eqref{eq:an} the value of $E_2$ calculated from the Schr\"odinger equation \eqref{eq:Schrodinger} with the potential $V^{(2)}(r)$. Somewhat surprisingly, such modified Eq.~\eqref{eq:an} still well describes the temporal positions of the vortices, so Eq.~\eqref{eq:positions1} remains a valid approximation. On the other hand, the values of the relative coordinate $r$ where  the vortices occur are underestimated by Eq.~\eqref{eq:r0} for $\lambda \gtrsim 1$, since the $\delta$-potential approximation does not take into account the actual quasi-rectangular shape of the potential $V^{(2)}(r)$. 
 
The first vortex in Eq.~\eqref{eq:an} occurs when
 $| E_{2}|R=3\pi/4$, that is, when  the bound and continuous spectrum terms acquire the opposite phases.
 Hence we can assume that the total phase in a finite medium of length $L$ required for at least one vortex-antivortex pair to form is given by $|E_{2}|L\ge 3\pi/4$.
Introducing the phase $\varphi =(\OD/2)(q\Gamma/\DeltaP)$ gained by a single photon in the medium, we obtain
from the condition $| E_{2}|L\ge3\pi/4$
that 
\begin{equation}\label{eq:phase}
\varphi\ge\frac{3 \sqrt{2\pi}}{8}\frac{\pi}{\lambda}=\frac{\varphi_0}{\lambda}\:.
\end{equation}
This is the condition used to plot an analytical curve in the phase diagram in Fig.~1(e) in the main text ($\varphi_0\approx 0.94\pi$).
Such an approximation qualitatively  describes also the positions of the vortices not only in the $\delta$-potential but also in the actual potential $V^{(2)}(r)$.

For large interaction strength $\lambda\gg 1$, we can still use the condition $| E_{2}|L=3\pi/4$ but the energy of the bound state should be estimated differently. In the crudest approximation, we can assume that the state is at the bottom of the well, $|E_{2}|=V^{(2)}(r=0)$.
In this case, we obtain the same equation as  Eq.~\eqref{eq:phase} but with $\lambda$ replaced by unity, \textit{i.e.}, $\varphi\ge\varphi_0$. This corresponds to the vertical line in the phase diagram in the main text. 

\section{Auxiliary experiments}
\subsection{Spectroscopy}
\begin{figure*}
\centering\includegraphics[width=0.45\textwidth]{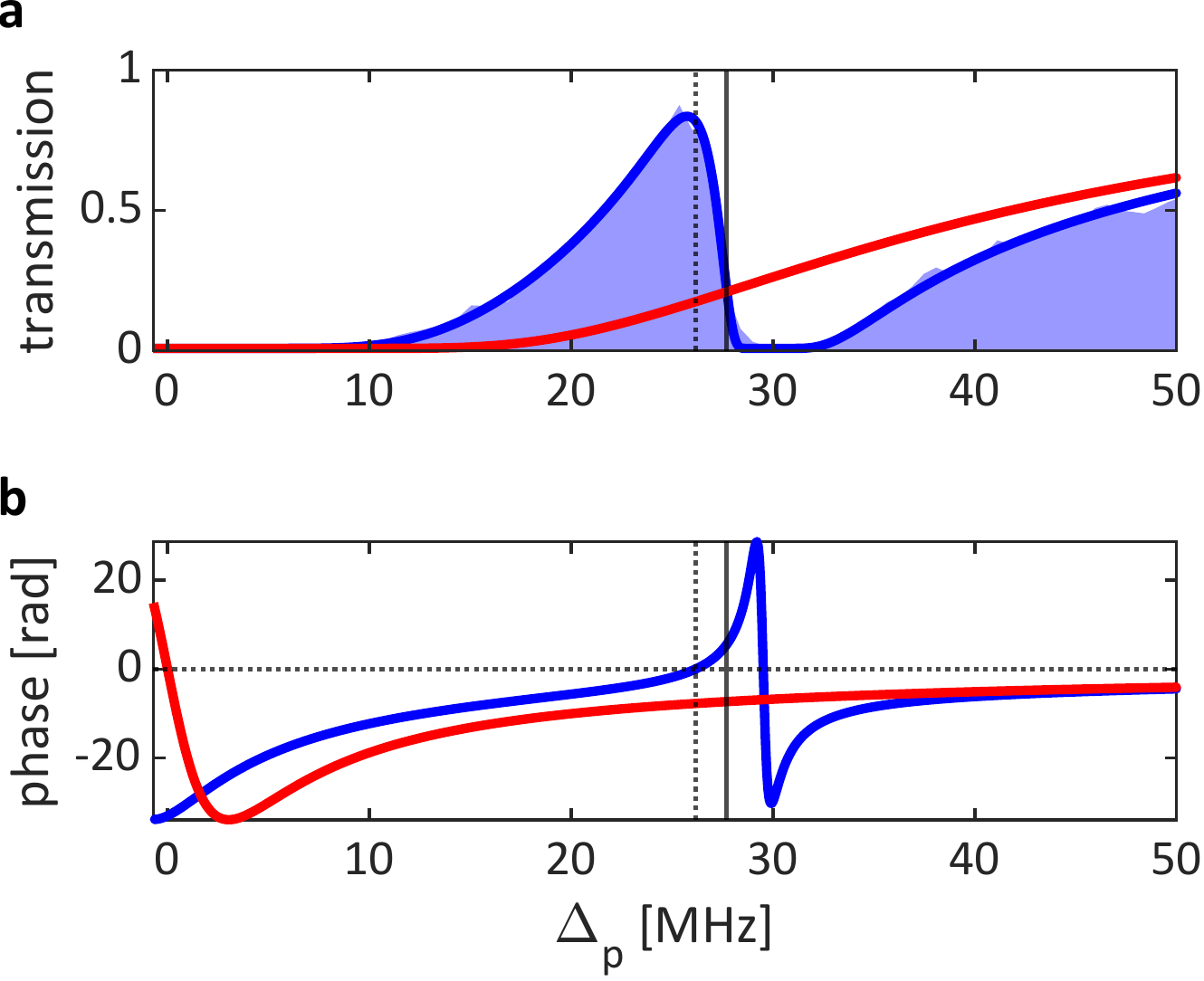}
	\caption{\textbf{Spectroscopy.} \textbf{a} Measured transmission spectroscopy (blue shaded area), fitted to a three-level model (blue line), from which a two-level transmission spectrum is inferred (red line). The EIT frequency is marked by a dashed vertical line, and the TE frequency (transmission equality, where the two and three-level transmissions are equal) is marked by a solid vertical line. 
 \textbf{b} Inferred single-photon phase for (blue) three-level and (red) two-level models.}
	\label{fig: spectra}
\end{figure*}

We use transmission spectroscopy measurements to find the system parameters ($\OD$, $\Omega$, $\Delta_\mathrm{c}$, $\gamma$) and to locate the transmission equality (TE) point. Specifically, we measure the transmission spectrum of the probe field in the presence of the control field as a function of OD and fit the spectra to an atomic three-level model. Figure \ref{fig: spectra}(a) shows an example of such measured spectrum (shaded area) and fit (blue line). The expected transmission in the absence of control is calculated from the fitted parameters using a two-level model. With the fit parameters at hand, we can calculate the expected transmission phase, as exemplified in Fig.~\ref{fig: spectra}(b). Close to the TE, highlighted by the solid vertical line, the two-level phase (red) is nearly constant. In contrast, the three-level phase (blue) rapidly changes, going from 0 at EIT to a value that is comparable (but opposite in sign) to the two-level phase at TE. This enhancement of the phase difference is quantified by $1\le q \le 2$.

\subsection{Dependence of \texorpdfstring{$\phi^{(2)}$}{measured phase} on probe power}
\begin{figure*}
	\centering
	\includegraphics[width=0.5\textwidth]{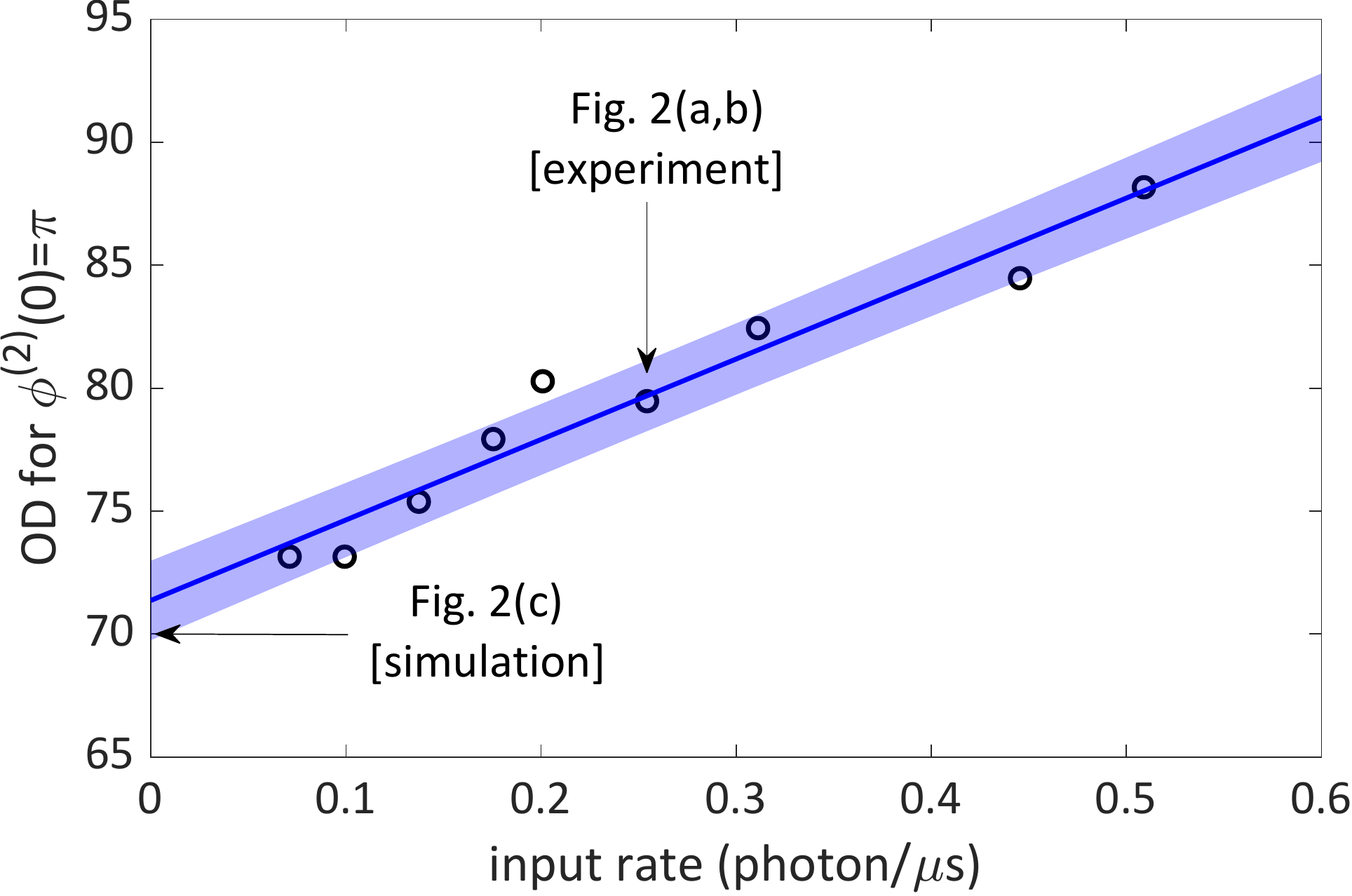}
	\caption{\textbf{Effect of probe power on the conditonal phase $\phi^{(2)}$}. The OD for which $\phi^{(2)}(0)=\pi$ is plotted for several different probe input rates. The data is fitted to a line (blue) with a slope of $33\pm3~\mu$s and an intercept of $71.4\pm0.9$. The $67\%$ fit confidence level is shown as a blue-shaded area. Arrows indicate the simulation and experimental results shown in Fig. 2 of the main text.
 }
	\label{fig: phaseVsPower}
\end{figure*}

As explained in Methods, the measurement of $\phi^{(2)}$ involves the measurement of the linear transmission phase $\phi^{(1)}$. In effect, $\phi^{(2)}(\tau)$ is extracted by subtracting $\phi^{(1)}$ from the phase of the conditional beat $B^{(1)}(t_\phi,\tau)$. To measure $\phi^{(1)}$, we detect the beat signal $B^{(1)}(t_\phi)$ without conditioning on detection in another detector. The fit to $B^{(1)}(t_\phi)=a+b\cos{[\nu t_\phi+\phi^{(1)}]}$, from which we extract $\phi^{(1)}$, then implicitly assumes that the number of photon pairs in the incoming (coherent state) probe field is negligible compared to the number of single photons. When this is not the case, the measured phase is altered by the photon-photon interaction, and our estimation of $\phi^{(1)}$ would be biased. As a result, we will underestimate $\phi^{(2)}$. 
To test the sensitivity of $\phi^{(2)}$ on the input photon rate, we have measured $\phi^{(2)}$ for several different rates and, for each rate, identified the OD for which $\phi^{(2)}(0)=\pi$, as shown in Fig.~\ref{fig: phaseVsPower}. These measurements explain the discrepancy between the $\OD\approx80$, for which vortices formed in the experiment shown in Fig.~2(a,b) in the main text, and the $\OD\approx70$, for which vortices form in the calculation in Fig.~2(c) (which simulates the ideal case of a vanishing input photon rate).

\bibliography{bibliography}